\documentclass[11pt,a4paper]{article} 
\usepackage{jcappub}
\usepackage{bm}
\usepackage{times} 
\usepackage[utf8]{inputenc}
\usepackage[english]{babel}
\usepackage[T1]{fontenc}
\usepackage{graphicx}
\usepackage{amsmath}
\usepackage{subfig}
\usepackage{hyperref}
\usepackage{color}
\usepackage{amssymb}
\usepackage[font=scriptsize,labelfont=bf]{caption}
\usepackage{appendix}
\usepackage{slashed}
\usepackage{slantsc}
\usepackage{braket}
\usepackage{mathrsfs}
\usepackage{subfig}
\usepackage{placeins}
\usepackage{fullpage}
\usepackage{sidecap}
\usepackage{float}
\usepackage{wrapfig}
\usepackage[dvipsnames]{xcolor}
\usepackage{dsfont}
\usepackage{framed}

\begin{document}

\begin{titlepage}

\setcounter{page}{1} \baselineskip=15.5pt \thispagestyle{empty}
\pagenumbering{roman}

\bigskip

\vspace{1cm}
\begin{center}
{\fontsize{20}{28}\selectfont \bfseries  Light-Cone Observables and Gauge-Invariance in the Geodesic Light-Cone Formalism}
\end{center}

\vspace{0.2cm}

\begin{center}
{\fontsize{13}{30}\selectfont Fulvio Scaccabarozzi$^a$ and Jaiyul Yoo$^{a,b}$}
\end{center}

\begin{center}
\vskip 8pt
\textsl{$^a$ Center for Theoretical Astrophysics and Cosmology,
Institute for Computational Science}\\
\textsl{University of Z\"urich, Winterthurerstrasse 190,
CH-8057, Z\"urich, Switzerland}

\vskip 7pt

\textsl{$^b$Physics Institute, University of Z\"urich,
Winterthurerstrasse 190, CH-8057, Z\"urich, Switzerland}
\end{center}

\note{fulvio@physik.uzh.ch,~~~~jyoo@physik.uzh.ch}

\vspace{1.2cm}
\hrule \vspace{0.3cm}
\noindent {\sffamily \bfseries Abstract} \\[0.1cm]
The remarkable properties of the geodesic light-cone (GLC) coordinates allow analytic expressions for the light-cone observables, providing a new non-perturbative way for calculating the effects of inhomogeneities in our Universe. However, the gauge-invariance of these expressions in the GLC formalism has not been shown explicitly. 
Here we provide this missing part of the GLC formalism by proving the gauge-invariance of the GLC expressions for the light-cone observables, such as the observed redshift, the luminosity distance, and the physical area and volume of the observed sources. Our study provides a new insight on the properties of the GLC coordinates and it complements the previous work by the GLC collaboration, leading to a comprehensive description of light propagation in the GLC representation. 
\vskip 10pt
\hrule

\vspace{0.6cm}
\end{titlepage}

\noindent\hrulefill \tableofcontents \noindent\hrulefill

\pagenumbering{arabic}

\section{Introduction}

The next generation of galaxy surveys will probe the Universe with high precision at very large scales \cite{Amendola:2012ys,Ivezic:2008fe,Hill:2008mv}. 
Due to the precision achieved by observations, the theoretical representation of what is observed can no longer rely on the assumption that our Universe is homogeneous and isotropic. Indeed, the light we measure in galaxy surveys is affected by the local inhomogeneities distributed along its path. To attain the level of accuracy set by the precision of observations, theoretical predictions must take into account the relativistic effects generated by the inhomogeneities, which play a key role at the large scales explored (see for instance \cite{Yoo:2013tc}). Only in this way we can avoid misinterpretation of surveys' measurements and extract the maximum physical information underlying the data (see \cite{Yoo:2010ni,Raccanelli:2015vla}). 

Many studies have been devoted to developing a relativistic description of the observables containing the information carried by the light. In most works, the description of the inhomogeneities in our Universe is obtained by adding perturbations to a homogeneous and isotropic FRW metric (see e.g. \cite{Bardeen}). In this case, the application of perturbation theory and general relativity enables the derivation of theoretical expressions for physical observables, accounting for the effects of the inhomogeneities to a certain perturbative level. Furthermore, it is often the case that specific gauge conditions are imposed to the metric perturbations before the calculations are performed. 

In order to simplify the task of making theoretical predictions in the context of general relativity, the geodesic light-cone (GLC) coordinates were introduced in \cite{GLC2011}. The GLC coordinates belong to a larger class of adapted coordinates that goes back to the pioneering works \cite{Temple,Kristian,Saunders 1,Saunders 2}. 
Contrary to the perturbative approach, the GLC coordinate system defines an exact (non-perturbative) metric representation of our Universe accounting for inhomogeneities. This representation is greatly helpful for problems associated with the observation of light sources lying on the past light-cone of a given observer, allowing fully non-linear and simple expressions of light-cone observables: the observed redshift, the luminosity distance, and the physical area and volume occupied by sources.
Once the expression of a given observable is obtained analytically in the GLC representation, it can be expressed perturbatively in any choice of gauge conditions by connecting the GLC metric to the chosen gauge with a coordinate transformation valid at the desired order in the perturbative theory. 

In \cite{GLC2012,BenDayan:2012,BenDayan:2013} the GLC metric was expressed in the conformal Newtonian gauge (or Poisson gauge when the calculations are extended to second order), computing the observed redshift and the luminosity distance in the presence of inhomogeneities. 
In these works, the GLC angular coordinate was intended to describe the observed angle of the source. However, the subtle difference between the observed angle in the observer rest frame and that in a global coordinate was neglected. Furthermore, the presence of additional degrees of freedom in the GLC variables was not taken into account. This was considered later in \cite{GLC2013,Fleury:2016}, but without describing explicitly how to make use of the residual gauge freedom associated to the GLC representation.

In \cite{GLC2013} the normalization condition for the GLC angular coordinate was fixed, bringing the expression of the luminosity distance derived with the GLC approach fully consistent with other approaches. Indeed, as we showed in \cite{Yoo:2016}, the geometric approach, the Sachs approach, the Jacobi mapping approach and the GLC approach reproduce the same correct prediction in the conformal Newtonian gauge.

After the correction suggested in \cite{GLC2013}, the GLC approach has been successfully used to calculate the expressions of the light-cone observables up to second order in perturbation theory in the Poisson gauge (see \cite{GLC2013,Marozzi:2014,DiDio,Fanizza:2015}). However, an explicit proof of gauge-invariance for these expressions is missing in the literature. 
According to the general covariance of general relativity, any coordinate can be used, but the expressions of observables must be the same in any choice of gauge conditions \cite{Bardeen}. The purpose of this work is to provide this missing part of the GLC formalism. 
Despite the consistency of the previous results \cite{GLC2012,GLC2013,BenDayan:2012,BenDayan:2013,Marozzi:2014,Fanizza:2015,DiDio}, we believe that it is important to prove the gauge-invariance by connecting the GLC metric to the most general perturbed FRW metric without choosing a gauge condition. This proof will ensure that the GLC expressions for the light-cone observables are identical in any gauge conditions beyond the gauge choices studied in previous works. In our derivation we will take into account perturbations to the first order, and we will consider all possible degrees freedom associated to the GLC variables, showing that the final expressions for the light-cone observables are independent from our normalization.
Furthermore, we will check the consistency with the approach introduced in \cite{Yoo:2009,Yoo:2010,Yoo} to describe the propagation of light in an inhomogeneous universe. This latter successfully reproduces the light-cone observables in a covariant and gauge-invariant way, providing us with a yardstick to compare all results. 

The organization of the paper is as follows. In sec.~\ref{GLC coord}, we introduce the GLC coordinates, describing their properties and features. In sec.~\ref{sec Gauge-invariance of the GLC observables}, we express the GLC variables and metric components in terms of the metric perturbations of a general FRW representation. In sec.~\ref{GLC variables}, we take a gauge transformation of the metric perturbations. Then, we calculate how the GLC quantities change accordingly. Then we derive with the GLC approach the expressions of the observed redshift in sec.~\ref{Redshift parameter}, the luminosity distance in sec.~\ref{Luminosity Distance}, and of the source volume in sec.~\ref{Galaxy Number Density}, showing the gauge-invariance. We conclude with a discussion in sec.~\ref{Conclusions}.

Throughout the paper, we set the speed of light $c \equiv 1$, we use the Greek indices $\mu,\nu,\rho,\sigma$ for the spacetime components, the Greek indices  $\alpha,\beta,\gamma,\delta$ for the spatial components and the Latin indices $a,b,c,d$ for the angular components.

\section{GLC representation}
\label{GLC representation}

Adopting the GLC representation one can write down exact (non-perturbative) expressions for light-cone observables. For this reason, it has been successfully used to derive the expressions for the observed redshift, the luminosity distance of faraway galaxies and the observed galaxy number density. However, in order to compute these expressions, one always has to convert the final expressions of these observables into those in the FRW metric with a particular choice of gauge conditions. 
Since physical quantities should be independent of our choice of gauge conditions for computation, this procedure should not cause any ambiguity, provided that the GLC approach is valid in any of these gauge choices. In this section, after presenting the GLC coordinates in detail, we perform a coordinate transformation from the GLC representation to the most general FRW metric representation at first order in perturbations. Then we take a gauge transformation and study how the GLC quantities transform. 

\subsection{GLC coordinates and their main properties}
\label{GLC coord}

The GLC coordinates, first introduced in \cite{GLC2011}, constitute a special coordinate system, which is particularly suitable when the purpose is to extract physical information from the light emitted by distant sources. 
The GLC coordinates $x^\mu_{\text{GLC}}=(w,\tau,\tilde{\theta}^a)$ are defined by the line element\footnote{See \cite{Fleury:2016} for the construction of the GLC line element through the coordinate basis vectors $\vec{\partial}_w,\vec{\partial}_\tau,\vec{\partial}_a$.} 
\begin{equation}
ds^2_{\text{GLC}} = \Upsilon^2dw^2 -2 \,\Upsilon \, dw \, d\tau + \gamma_{ab}\,(d\tilde\theta^a-U^a dw)(d\tilde\theta^b-U^b dw)\,, 
\end{equation}
which specifies the metric tensor in the GLC representation: 
\begin{equation}
\label{GLC metric}
\begin{split}
g_{\mu\nu}^{\text{GLC}}=
\begin{bmatrix}
\,\Upsilon^2+U^2 && -\Upsilon\, && -U_b \,
\\\\
-\Upsilon && 0 && \vec{0}
\\\\
-U_a && \vec{0} && \gamma_{ab}
\end{bmatrix},
\qquad\qquad
g^{\mu\nu}_{\text{GLC}}=
\begin{bmatrix}
0 & -1/\Upsilon & \vec{0}
\\\\
-1/\Upsilon & -1 & -U^b/\Upsilon
\\\\
\vec{0} & -U^a/\Upsilon & \gamma^{ab}
\end{bmatrix}   \,,
\end{split}
\end{equation}
\[
\sqrt{-g} = \Upsilon\sqrt{\vert \gamma \vert}\,,\qquad g = \text{det}\,g_{\mu\nu} \,,\qquad \gamma = \text{det}\,\gamma_{ab} \,, \qquad \mu,\nu = w,\tau,\tilde\theta,\tilde\phi, \qquad a,b=\tilde\theta,\tilde\phi\,.
\]
In such coordinates, a generic space-time point is identified by a past light-cone hypersurface $w$, a proper-time hypersurface $\tau$, and the angular position $\tilde{\theta}^a$ measured by the observer at the tip of the light-cone. In accordance with this definition, $w$ generates the photon wavevector $k_\mu=\partial_\mu w$ and is therefore a null coordinate ($\partial^\mu w \,\partial_\mu w=0$), $\tau$ generates the observer four-velocity $u_\mu=-\partial_\mu \tau$, which follows a geodesic flow ($u^\nu\nabla_\nu u^\mu=0$) and satisfies $k^\mu u_\mu=\Upsilon^{-1}$, while $\tilde{\theta}^a$ parametrizes a two-sphere orthogonal to the photon wavevector and is constant along the null geodesics ($k^\mu\partial_\mu\tilde{\theta}^a=\vec{0}\,$). As we shall see, the metric components can be interpreted as follows: $\Upsilon$ describes the expansion of the universe, $\gamma_{ab}$ is the induced metric on the two-sphere of constant time, $U^a$ represents a measure of the space-time anisotropy \cite{BenDayan:2012}. 

The physical meaning of the GLC variables and metric components becomes evident when we consider a homogeneous universe. For a spatially homogeneous and isotropic FRW metric
\begin{equation}
 ds^2=a^2(\eta)(-d\eta^2 + d r^2+ r^2d \theta^2+ r^2\sin^2 \theta d\phi^2)\,,
\end{equation}
the transformations from a GLC coordinate $x^\mu_{\text{GLC}}=(w,\tau,\tilde\theta,\tilde\phi\,)$ and metric components $g^{\mu\nu}_{\text{GLC}}$ to a FRW coordinate $y^\mu_{\text{FRW}}=(\eta,r,\theta,\phi)$ are given by
\begin{equation}
\label{BG relation}
\begin{split}
w &=\eta +  r \,, \qquad\qquad \tau= t \,, \qquad\qquad \tilde{\theta}^a= \theta^a =(\theta,\phi)\,,
\\
\Upsilon &=a \,, \qquad\qquad\quad U^a=0\,, \,\,\,\quad\quad\quad \gamma_{ab}=a^2 \, \bar g_{ab} \,,
\end{split}
\end{equation}
where $a(\eta)$ is the expansion scale factor, $\eta$ is the conformal time, $t$ is the proper-time (such that $dt=a\,d\eta$), $\bar g_{ab}=\text{diag}(r^2,r^2\sin^2\theta)$, which lowers the two-dimensional indices, and the FRW metric is written in spherical coordinates. Mind the difference of the GLC angles $(\tilde\theta,\tilde\phi)$ and the FRW coordinates $(\theta,\phi)$. For future use, we also define the two-dimensional (angular) tensor $\hat{\bar{g}}_{ab}=\text{diag}(1,\sin^2\theta)$, so that $\bar g_{ab}=r^2\, \hat{\bar{g}}_{ab}$.

When inhomogeneities in our Universe are taken into account, the light-cone hypersurface $w$ and its intersection with the proper-time hypersurface $\tau$ are no longer a cone and a two-sphere, as inhomogeneities generate geometric distortions. However, when no caustics form on the past light-cone, these inhomogeneous surfaces are still \textit{topologically equivalent} to a cone and a uniform two-sphere.\footnote{Geometric distortions of the light-cone hypersurface may lead to the intersection of light rays, at points called caustics. In this situation the GLC formalism fails, as the topological equivalence with a cone and a uniform sphere obviously breaks down. This issue becomes important for instance at small scales where strong lensing is involved.} Consequently, in the GLC representation, photons travel along the straight line connecting the source point on the topological two-sphere and the tip of the topological cone. This straightforward geometry represents the great advantage of the GLC representation, leading to the simple expressions of light-cone observables. 

\subsection{Coordinate transformation from GLC to perturbed FRW} 
\label{sec Gauge-invariance of the GLC observables}

Physical observables must be the same in any choice of gauge conditions, regardless of the method adopted for the derivation. 
Our goal is to establish the gauge-invariance of light-cone observables derived in the GLC approach.
So far light-cone observables in the GLC representation have been expressed in the conformal Newtonian gauge and in the synchronous gauge. Despite the consistency of the previous results \cite{GLC2011,GLC2012,GLC2013,BenDayan:2012ct,Fanizza:2014baa,BenDayan:2012,BenDayan:2013,Marozzi:2014,Fleury:2016,Fanizza:2015,Nugier:2015awa,DiDio}, we believe that it is important to prove the gauge-invariance by adopting the most general metric without choosing a gauge condition. This proof will ensure that the GLC expressions for the light-cone observables are identical in any gauge conditions beyond the two gauge choices studied in previous works.

First of all, we need to take a coordinate transformation from the GLC to the fully general perturbed FRW representation accounting for inhomogeneities. In this representation the description of the physical universe is obtained by adding perturbations to a homogeneous and isotropic FRW metric. Considering perturbations only to first order, the most general perturbed FRW metric tensor describing the physical universe is
\begin{equation}
\label{most general metric}
\begin{split}	
	g_{\mu\nu}^{\text{FRW}}=
	a^2\begin{bmatrix}
		-(1+2\mathcal A) & & & -\mathcal B_{\alpha}&
		\\\\
		-\mathcal B_{\alpha} 	& & &  (\,\bar{g}_{\alpha\beta}+2\,\mathcal C_{\alpha\beta})&
		\\
		& &	& &
	\end{bmatrix},
\quad
g^{\mu\nu}_{\text{FRW}}=\frac{1}{a^2}\begin{bmatrix}
		-(1-2\mathcal A)& & -\mathcal B^\alpha
		\\\\
		-\mathcal B^\alpha &	&  (\,\bar{g}^{\alpha\beta}-2\,\mathcal C^{\alpha\beta}\,) \,
\\
&			
			\end{bmatrix},  
\end{split}
\end{equation}
where $\alpha,\beta = r,\theta,\phi,$ and the small perturbations from the background metric are represented by\footnote{The notation $C_{(\alpha|\beta)}$ means symmetrization of the indices. Analogously $C_{[\alpha|\beta]}$ means antisymmetrization.}$^,$\footnote{By constructing the line element from the metric tensor in eq.~\eqref{most general metric}, the dimensions of the perturbations are
$
[\alpha]=[\varphi]=1\,, \,\, [\beta]=L\,, \,\, [\gamma]=L^2 \,,
\,\,
[B_r]=1\,, \,\, [B_a]=L \,, 
\,\,
[C_r]=L \,, \,\, [C_a]=L^2 \,,
\,\,
[C_{rr}]=1 \,, \,\, [C_{ra}]=L \,, \,\, [C_{ab}]=L^2 \,, 
$ where $L$ is the dimension of a length.}
\begin{equation}
\label{metric perturbations}
	\begin{split}
		\delta g_{00}&\equiv -2 \,a^2 \mathcal{A}\equiv -2\, a^2 \alpha\,, \qquad \quad \delta g_{0\alpha}\equiv -a^2 \mathcal{B}_\alpha \equiv -a^2 (\beta_{,\alpha}+B_{\alpha})\,,
		\\
		\delta g_{\alpha\beta}&\equiv 2 \, a^2 \mathcal{C}_{\alpha\beta}\equiv 2a^2 (\varphi \, \bar{g}_{\alpha\beta}+\gamma_{,\alpha | \beta}+C_{(\alpha|\beta)}+C_{\alpha\beta})\,.
	\end{split}
\end{equation}
We decomposed the metric perturbations into scalars ($\alpha,\beta,\varphi,\gamma$), vectors ($B_\alpha,C_\alpha$) and tensors ($C_{\alpha\beta}$), where the vector perturbations are divergenceless and the tensor perturbations are both divergenceless and traceless:
\begin{equation}
	B^\alpha_{\,\,\,\,|\alpha}=0 \,, \qquad\qquad C^\alpha_{\,\,\,\,|\alpha}=0 \,, \qquad\qquad C^{\alpha\beta}_{\,\,\,\,\,\,\,|\alpha}=0 \,,\qquad\qquad C^{\alpha}_{\alpha}=0 \,.
\end{equation}
Here the perturbations depend on the space-time point, the comma is the ordinary derivative and the vertical bar is the covariant derivative with respect to the three-spatial metric $\bar{g}_{\alpha\beta}$, which lowers the three-dimensional indices (the affine connections are given in appendix \ref{appendix A}). 
In a global coordinate $y^\mu_{\text{FRW}}=(\eta,r,\theta,\phi)$, the three-spatial metric is $\bar{g}_{\alpha\beta}=\text{diag}(1,r^2,r^2\sin^2\theta)$ and a space-time point is identified by a conformal time and spherical coordinates with origin at the position of an observer moving with time-like four-velocity $u^\mu \equiv a^{-1}(1-\alpha\,,\,V^\alpha)$.

The GLC metric tensor in eq.~\eqref{GLC metric} is related to the FRW metric tensor in eq.~\eqref{most general metric} through a coordinate transformation from $x^\mu_{\text{GLC}}=(w,\tau,\tilde\theta^a)$ to $y^\mu_{\text{FRW}}=(\eta,r,\theta^a)$:
\begin{equation}
\label{GLC -> MG}
	g_{\text{GLC}}^{\mu\nu}(x)=\frac{\partial x^\mu}{\partial y^\rho}\frac{\partial x^\nu}{\partial y^\sigma}g^{\rho\sigma}_{\text{FRW}}(y)\,.
\end{equation}
By solving these differential equations, we obtain the perturbative form of the GLC quantities. As a result, the coordinates $w,\,\tau,\,\tilde\theta^a$ and the functions $\Upsilon,\,U^a,\,\gamma^{ab}$ will be expressed in terms of the coordinates $\eta,\,r,\,\theta^a$ and the metric perturbations $\mathcal A,\,\mathcal B_\alpha,\,\mathcal C_{\alpha\beta}$.
In order to solve the differential equations, we first split the GLC variables into the background and perturbation quantities:
$w=\bar{w}+\delta w \,,\,\, 
\tau = \bar \tau + \delta \tau \,, \,\,
\tilde\theta^a = \bar{\tilde\theta}^a + \delta\tilde\theta^a$. In this way we can simplify the calculations by making use of the background relations in eq.~\eqref{BG relation}.
Furthermore, we parametrize the background path $\bar x^\mu(\bar r)=(\bar\eta_o-\bar r,\bar r,\theta^a)$ of a photon traveling from a source to the observer with an affine parameter $\bar r$ representing the comoving distance 
\begin{equation}
\label{bg photon path}
	 \bar r=  \bar \eta_o - \eta = \int_0^{\bar z(\eta)} \frac{dz}{H(z)}\,,
\qquad\qquad\qquad
	1+\bar z (\eta) \equiv \frac{a(\bar \eta_o)}{a(\eta)}	\,.
\end{equation}
Here $\bar \eta_o$ is the conformal time of the observer today in a homogeneous universe, $H(z)$ is the Hubble parameter, and $1+\bar z (\eta)$ is the redshift parameter of a time coordinate $\eta$.\footnote{In cosmology, the observed redshift $z$ provides the unique physically meaningful way to express the time coordinate of a source. In the presence of inhomogeneities, the observed redshift $z$ is split into the background expansion contribution $\bar z$ and a perturbation $\delta z$, such that $1+z \equiv (1+\bar z)(1+\delta z)$. The observed redshift is used to infer the source coordinate time $\bar \eta_z$ using the distance\,-\,redshift relation in a homogeneous universe
\[
\bar \eta_o - \bar\eta_z = \int_0^{z} \frac{dz}{H(z)} \,,
\]
and the coordinate time $\bar\eta_z$ associated with the observed redshift is different from the source coordinate time $\eta$ associated with the redshift parameter $\bar z$ (see eq.~\eqref{gi redshift}). 
Note that the conformal time today in a homogeneous universe is uniquely determined (given a set of cosmological parameters) as
$
\bar \eta_o = \int_0^{\infty} dz/ H(z)\,,
$
and the scale factor $a$ is usually set to unity at $\bar\eta_o$.} The tangent vector to the unperturbed photon geodesic $\bar x^\mu(\bar r)$ is the background photon wavevector $\bar k^\mu$ and the variation of a given function $f$ along $\bar x^\mu(\bar r)$ is given by 
\begin{equation}
\frac{df}{d\bar r}=\frac{d \bar x^\mu}{d\bar r}\frac{\partial f}{\partial \bar x^\mu} =\bar k^\mu\partial_\mu\, f=-\frac{\partial f}{\partial \eta} + \frac{\partial f}{\partial r}\,, \qquad\qquad \bar k^\mu = \frac{d\bar x^\mu}{d\bar r} \,.	
\end{equation}
In \cite{GLC2012} the light-cone variables $\eta_\pm \equiv \eta \pm r$ and the corresponding partial derivatives $\partial_\pm = (\partial_\eta \pm \partial_r)/2$, were introduced to simplify the coordinate transformation in eq.~\eqref{GLC -> MG}. The conversion between the light-cone variables and our (background) affine parameter $\bar r$ is given by
\begin{equation}
	\partial_- = -\frac{1}{2}\frac{d}{d\bar r} \,,
		\qquad\qquad
		\partial_+ = \frac{\partial}{\partial\eta} + \frac{1}{2} \frac{d}{d\bar r} \,,
		\qquad\qquad
		\int_{\eta_+}^{\eta_-} d\eta'_-  = -2 \int_0^{\bar r} d\bar r' \,.
\end{equation}
Moreover, given a generic function $f(\bar r)$ integrated along the background photon path from the observer to a source, we can extract boundary terms in the following way:
\begin{equation}
\label{bgpp}
	\int_0^{\bar r_s}d\bar r \, \partial_r f = \int_0^{\bar r_s}d\bar r \, {f}' + f \big\vert^s_o \,,
\end{equation}
where a prime means the derivative with respect to conformal time, and $\bar r_s$ represents the value of the affine parameter $\bar r$ corresponding to the source point along the unperturbed photon geodesic. The integration over the comoving distance $\bar r$ can be directly translated into an integral over conformal time $\eta$, justifying the change of derivation for the integrands. The letters $``s\,"$ and  $``o\,"$ are used to represent that the quantities are evaluated at the source and observer positions, respectively.

Let us now put everything together to express the GLC coordinates in terms of the metric perturbations.
First, to obtain $w$ we consider the component $ww$ of eq.~\eqref{GLC -> MG}:
\begin{equation}
\label{diff eq w}
	\bar w = \eta + r \,, \qquad\qquad\qquad \frac{d}{d\bar r} \delta w=- \mathcal A + \mathcal B^r + \mathcal C^{rr}  \,.
\end{equation}
The solution of the differential equation can be written as
\begin{equation}
\label{sol w}
\delta w_s - \delta w_{o} =- \int_0^{\bar r_s} d\bar r \big[\mathcal A - \mathcal B^r - \mathcal C^{rr}\big]\,,
\end{equation}
where the integrand is a function of the position along the photon path $\bar r$. 
By using eq.~\eqref{bgpp} we extract the boundary terms and derive
\begin{equation}
\label{omega}
\begin{split}
w_s =  \eta_s + r_s &+\delta w_o -\int_0^{\bar r_s} d\bar r\,[\alpha-(\varphi+\beta'+\gamma''+B^r+{C^r}'+C^{rr})]
\\
& 
+ \big[\beta+\gamma'+\gamma^{,r}+C^r \big]^{s}_{o} \,.
\end{split}
\end{equation}
In the literature the integration constant $\delta w_{o}$ is often set to zero. At this point we do not specify the value of this quantity, which is related to the perturbations to the photon propagation at observation through the exact relation $k_\mu=\partial_\mu w$ (see below and appendix \ref{Identification of the GLC angles with the observed angles}).\footnote{It is worth noting that eq.~\eqref{omega} can also be obtained from the null condition $k^\mu k_\mu=g^{\mu\nu}\partial_\mu w \, \partial_\nu w=0$.}  
For a proper-time $\tau$ we consider the component $\tau\tau$ of eq.~\eqref{GLC -> MG}:
\begin{equation}
\label{tau}
\bar\tau = t \,, \qquad\qquad\qquad \frac{\partial}{\partial \eta}\delta\tau = a\,\alpha \,;
	\qquad\qquad\qquad
\tau_s = \int_{0}^{\eta_s}d\eta\, a\,[1+\alpha] \,.	
\end{equation}
For the GLC angles $\tilde\theta^a$ we consider the component $wa$ of eq.~\eqref{GLC -> MG}:
\begin{equation}
\label{diff theta}
\begin{split}
	\bar{\tilde\theta}^{a} &=\theta^a \,, \qquad\qquad\qquad \frac{d}{d\bar r}  \delta\tilde\theta^{a} =  \mathcal B^a+2 \mathcal C^{ra} - \bar g^{ac} \partial_c \, \delta w  \,,
\\
\tilde\theta^a_s &=  \theta^a_s + \delta\tilde\theta^a_o + \int_0^{\bar r_s}d\bar r\, [\mathcal B^a+2 \mathcal C^{ra} - \bar g^{ac} \partial_c \, \delta w]  \,.
\end{split}
\end{equation}
We make use of eq.~\eqref{bgpp} to simplify the integration as
\begin{equation}
\label{theta}
\begin{split}
\tilde\theta^a_s  =  \theta^a_s & + \delta\tilde\theta^a_o -\bar r_s \, \delta w^{,a}_o  + \bar r_s \big[(\beta+\gamma')^{,a} + (\gamma^{,r}+C^r)^{,a} \big]_o 
\\
&  + \int_0^{\bar r_s} d\bar r \bigg(\frac{\bar r_s -\bar r}{\bar r_s \bar r}\bigg) \hat{\bar{g}}^{ac}\partial_c [\alpha-(\varphi+\beta'+\gamma''+B^r+{C^{r}}'+C^{rr})]
		\\
		&	+ \int_0^{\bar r_s}d\bar r \, \big[B^a+{C^{a}}'+2\,C^{ra} \big] + \big[ \gamma^{,a} + C^a \big]^s_o 
		 \,.
\end{split}
\end{equation}
The quantities $\delta\tilde\theta^a_o$ and $\delta w_o$ represent initial conditions, related to our choice of normalization at the observer point. 
These degrees of freedom are related to the residual gauge freedom of the GLC representation pointed out in \cite{GLC2013,Fleury:2016}. Indeed, as described in \cite{Fleury:2016}, the definition of the GLC coordinates in sec.~\ref{GLC coord} does not fully specify the choice of light-cone and its observed angles. Consequently, it is always possible to find coordinate transformations that redefine $w$ and $\tilde\theta^a$, but  leave the GLC metric unchanged.
These degrees of freedom should be set according to physical constraints, considering the observer peculiar velocity, the gravitational potential at the observer position and the orientation of directions in the observer rest frame with respect to the global coordinates.
In appendix \ref{Identification of the GLC angles with the observed angles} we show how to properly fix such freedom in order to match the GLC angles with the angles measured by the observer in the rest frame (the observed angles).
However, any different choice can be made (for instance, one can set $\delta\tilde\theta^a_o=\delta w^{,a}_o = 0$) with the GLC angles corresponding to the observed angles plus a constant at the observer. Naturally, the final expression of physical observables should not depend on our parametrization, as we show in sec.~\ref{Luminosity Distance}, where we derive the luminosity distance without choosing any particular normalization for $\delta\tilde\theta^a_o$ and $\delta w^{,a}_o$. 

Now, starting again from eq.~\eqref{GLC -> MG}, we derive the remaining components of the GLC variables in terms of the metric perturbations.
First, for the expansion factor $\Upsilon$ we simply consider the component $w\tau$ of eq.~\eqref{GLC -> MG}:
\begin{equation}
\label{UPSILON}
\Upsilon = a(\eta)\big[ 1 +\alpha -V^r - \delta w'\, \big] \equiv a\,[1+\delta\Upsilon] \,,
\end{equation}
where we defined the fractional perturbation $\delta\Upsilon$.
The induced metric $\gamma^{ab}$ is obtained by considering the component $ab$ of eq.~\eqref{GLC -> MG}:
\begin{equation}
\label{GAMMA}
		\gamma^{ab} = \frac{1}{a^2}\big\lbrace (1 - 2\,\varphi)\bar g^{ab}+\big[ \bar g^{ac}\partial_c\, \delta\tilde\theta^b -(\gamma^{,a|b}+C^{a|b}+C^{ab} )
		+ a\leftrightarrow b
		\big]
  \big\rbrace \equiv \frac{1}{a^2}[\,\bar g^{ab}+\delta \gamma^{ab}\,] \,,
\end{equation}
where we also defined the fractional fluctuation $\delta\gamma^{ab}$.
Finally, to derive the solution for $U^a$ we need to consider the component $\tau a$ of \eqref{GLC -> MG}:
\begin{equation}
\label{U}
U^ a = V^a + {\delta\tilde\theta^a}' \,.
\end{equation}
Since $U^a=0$ in the homogeneous background, there is no reason to define $\delta U^a$.
Note that the components of the peculiar velocity appearing in eqs.~\eqref{UPSILON} and \eqref{U} are given by
\begin{equation}
\label{velocity}
V^\alpha=\mathcal B^\alpha-\frac{1}{a} \,  \delta\tau^{,\alpha} \,,
\end{equation} 
which is obtained from considering the relation $u_\mu = -\partial_\mu \tau$, where $ u_\mu = g_{\mu\nu}u^\nu = -a\,(1+\alpha\,,\mathcal B_\alpha - V_\alpha)$.

\subsection{Gauge transformation of GLC variables}
\label{GLC variables}

In the previous section we expressed the GLC quantities in terms of the perturbations of a general metric representation. Using the gauge transformation of the metric perturbations we derive the gauge transformation of the GLC variables.

We consider the most general coordinate transformation:
$\hat{x}^\mu=x^\mu+\xi^\mu\,$, where $\xi^{\mu}=(T,\mathcal{L}^\alpha)$ and $\mathcal{L}^\alpha \equiv L^{,\alpha}+L^\alpha\,$. The transformations of the metric perturbations are well-known
\begin{equation}
\label{gauge transformation}
\begin{split}
\hat{\alpha} &= \alpha-T'-\mathcal{H}T \,, \quad\quad \hat{\beta}=\beta-T+L' \,, \quad\quad \hat{\varphi}=\varphi-\mathcal{H}T \,, \quad\quad \hat{\gamma}=\gamma-L \,,
\\
\hat{B}^\alpha &= B^\alpha+{L^\alpha}' \,, \quad\quad \hat{C}^\alpha=C^\alpha-L^\alpha \,,
\quad\quad \hat V^\alpha = V^\alpha+{\mathcal L^{\alpha}}' \,, \quad\quad \hat{C}^{\alpha\beta}=C^{\alpha\beta} \,,
\end{split}
\end{equation}
where $\mathcal H = a'/a = aH$ is the conformal Hubble parameter. 
Based on these gauge transformation properties we can define gauge-invariant quantities at linear level:
\begin{equation}
\label{GI variables}
	\alpha_\chi = \alpha - \frac{1}{a}\chi'\,, \quad \varphi_\chi = \varphi-H\chi \,, \quad \Psi^\alpha = B^\alpha + {C^\alpha}' \,, \quad \mathcal V^\alpha = V^\alpha + {\mathcal G^\alpha}' \,,
\end{equation}
where $\chi=a\,(\beta+\gamma')$ is the scalar shear of the normal observer, transforming as $\hat\chi=\chi-aT$. 
The notation for scalar gauge-invariant variables is set up such that $\alpha_\chi$ and $\varphi_\chi$ correspond to the gravitational potentials $\alpha$ and $\varphi$ in the conformal Newtonian gauge (where $\chi=0$) \cite{Yoo}.
In the same spirit, we defined $\mathcal G^\alpha = \gamma^{,\alpha}+C^\alpha$, which conversely is a pure gauge term transforming as $\hat{\mathcal G}^\alpha=\mathcal G^\alpha - \mathcal L^\alpha$. 
With these definitions we can rewrite the GLC quantities as
\begin{eqnarray}
\label{omega GI}
w_s &=& \eta_s+r_s +\delta w_{o} -\int_0^{\bar r_s} d\bar r\,[\alpha_\chi-\varphi_\chi-\Psi^r-C^{rr}]
+ \bigg[\frac{\chi}{a}+\mathcal G^r \bigg]^{s}_{o} \,,
\\
\label{tau GI}
	\tau_s &=& \int_{0}^{\eta_s}d\eta\, \big[ a\, (1+\alpha_\chi)  + \chi'\,\big]\,,
\\
\label{theta GI}
\tilde\theta^a_s &=&  \theta^a_s + \delta\tilde\theta^a_o  - \bar r_s \, \delta w^{,a}_o + \bar r_s \bigg[ \frac{\chi}{a}^{,a} + \mathcal G^{r,a} \bigg]_o + \mathcal G^a \big\vert^s_o  
\\\notag
 &&\quad \,+ \int_0^{\bar r_s} d\bar r \bigg[ \Psi^a+2\,C^{ra} + \bigg(\frac{\bar r_s-\bar r}{\bar r_s \bar r}\bigg) \hat{\bar{g}}^{ac}\partial_c (\alpha_\chi-\varphi_\chi-\Psi^r-C^{rr}) \bigg] \,,
\\
\label{UPSILON GI}
\Upsilon &=& a(\eta)\bigg[ 1 +\alpha_{\chi} - \mathcal V^r +H \chi  + \int_0^{ \bar r_s}d \bar r \, \big[\alpha_\chi-\varphi_\chi - \Psi^r - C^{rr} \big]'  \,\bigg]\,,
\\
\label{GAMMA GI}
\gamma^{ab} &=&  \frac{1}{a^2}\big\lbrace (1 - 2\,\varphi_\chi - 2\,H\chi\,)\,\bar g^{ab}+\big[\bar g^{ac}\partial_c\, \delta\tilde\theta^b -(\mathcal G^{a|b}+C^{ab} )
		+ a\leftrightarrow b
		\big]
  \big\rbrace \,,
\\
\label{U GI}
	U^a &=& \mathcal V^a - {\mathcal G^\alpha}' +  {\delta\tilde\theta^a}'  \,.
\end{eqnarray}
Thanks to the relations in eq.~\eqref{gauge transformation} we can derive how the GLC variables change under the gauge transformation:
\begin{equation}
	\delta\hat w_s   =\delta w_s + (\delta \hat w_{o} - \delta w_{o}) - \big[ T   +  \mathcal L^r \big]^s_o \,,
\end{equation} 
\begin{equation}
	\delta\hat\tau_s = \delta\tau_s  -a(\eta_s)\,T_s \,,
\end{equation}
\begin{equation}
\label{delta theta GT}
\delta\hat{\tilde{\theta}}^a_s  =\delta\tilde\theta^a_s + (\delta\hat{\tilde\theta}^a_o - \delta\tilde\theta^a_o) -  \bar r_s \,(\delta \hat w_o^{,a}-\delta w_o^{,a}) -  \bar r_s \,  \big [ T^{,a} + \mathcal L^{r,a} \big]_o - \mathcal L^a  \big\vert^s_o  \,,
\end{equation}
\begin{equation}
\label{gauge Upsilon}
	\delta\hat\Upsilon  =\delta\Upsilon -\mathcal H\, T \,,
\end{equation}
\begin{equation}
	\delta\hat\gamma^{ab}  = \delta\gamma^{ab} + 2\,\mathcal H T\, \bar g^{ab} +  \big[ (\delta\hat{\tilde\theta}^a - \delta\tilde\theta^a)^{,b} + \mathcal L^{a|b}   + a \leftrightarrow b  \big] \,,
\end{equation}
\begin{equation}
\label{gauge U}
		\hat U^a = U^a   \,.
\end{equation}
Clearly, the proper-time $\tau$ is a gauge-invariant physical observable. However, according to the way we split it, both the background part $\bar\tau$ and the perturbation $\delta\tau$ are gauge-dependent quantities,  and the gauge modes associated with the two parts cancel each other. The same argument applies to the GLC angles $\tilde\theta^a$ when the degrees of freedom in $\delta\tilde\theta^a_o$ and $\delta w^{,a}_o$ are set to match the observed angles in the rest frame of the observer (see appendix \ref{Identification of the GLC angles with the observed angles}), which are gauge-invariant physical observables.

\section{Gauge-invariance of the light-cone observables in GLC}
\label{3}

The position of a source galaxy is identified by the observed redshift $z_s$ and the observed angles $\theta^a_{\text{obs}}=(\theta_{\text{obs}},\phi_{\text{obs}})$, measured in the observer rest frame. Based on these quantities, the observer infers the source position $\bar x^\alpha$ by using the distance\,-\,redshift relation in a homogeneous universe. However, the real position $x^\alpha_s$ of the source is different from the inferred one $\bar x^\alpha_s$, because the inhomogeneities affect the photon propagation. To account for the effect of the inhomogeneities on the real source position with respect to the inferred position we define the distortion $\delta z$ in the observed redshift (related to the time distortion $\Delta\eta$) and the geometric distortions $(\delta r,\delta\theta,\delta\phi)$ of the source position. These can be computed by tracing the photon path backward from the observer to the source and solving for the real position, as described in \cite{Yoo}. On the other hand, the advantage of the GLC approach is that the distortions due to inhomogeneities are already incorporated in the coordinate system. As a consequence, the expressions of the light-cone observables in the GLC approach can be derived analytically.
In this section we derive the light-cone observables in the GLC approach and show that their final expressions are gauge-invariant.

\subsection{Observed redshift}
\label{Redshift parameter}

In GLC coordinates, the null geodesic connecting source and observer is characterized by the tangent vector $k^\mu= \delta^{\mu}_\tau \,\Upsilon^{-1}$, so that the coordinates $w$ and $\tilde\theta^a$ are constant along the photon propagation. Consider photons emitted by a geodesic source at the two-sphere identified by the past light-cone $w$ of the geodesic observer and the proper-time of emission $\tau_s$, and received by the observer at $\tau_o$. The observed redshift $z_s$ of these photons is then given by the exact relation \cite{GLC2011}
\begin{equation}
\label{z_s}
1+z_s=\frac{(k^\mu u_\mu)_s}{(k^\nu u_\nu)_o}=\frac{(\partial^\mu w\,\partial_\mu\tau)_s}{(\partial^\nu w\,\partial_\nu\tau)_o}=\frac{\Upsilon_o}{\Upsilon_s} \,.
\end{equation}
As a consequence, by using eq.~\eqref{UPSILON} and considering that the source is located on the observer past light-cone (given by $w=\eta_o$) at distance $r_s=\eta_o-\eta_s$, we obtain
\begin{equation}
\begin{split}	
\label{1+z}
		& 1+z_s = \frac{a(\eta_o)}{a(\eta_s)}\big[ 1+\delta\Upsilon_o-\delta\Upsilon_s \big] \,,
\\
	\delta\Upsilon_o = \alpha_{\chi o} - \mathcal V^r_o +H_o\chi_o  \,, \quad\quad &\delta\Upsilon_s = \alpha_{\chi s} - \mathcal V^r_s +H_s\chi_s  + \int_0^{ \bar r_s}d \bar r \, \big[\alpha_\chi-\varphi_\chi - \Psi^r - C^{rr} \big]' \,.
\end{split}
\end{equation}
In agreement with eq.~\eqref{gauge Upsilon}, these first order quantities gauge-transform as $\delta\hat\Upsilon_o = \delta\Upsilon_o  - \mathcal H_o T_o$ and $\delta\hat\Upsilon_s = \delta\Upsilon_s  - \mathcal H_s T_s$.

Before we proceed we need to consider a coordinate lapse, often ignored in literature (see \cite{Yoo:2016}): the observer time coordinate in an inhomogeneous universe deviates from its background value $\bar \eta_o$ by
\begin{equation}
\label{coord time effect}
	\delta\eta_o = - \frac{1}{a(\bar\eta_o)} \int_0^{\bar\eta_o} d\bar\eta \, a \, \alpha = - \frac{1}{a(\bar\eta_o)}\delta\tau_o  \,.
\end{equation}
This quantity represents the lapse between the coordinate time $\eta_o$ at the observer and the observer's proper-time $\tau_o$.\footnote{By considering the time component of the observer four-velocity $u^\mu=dx^\mu/d\tau$ we get the relation between the proper-time $\tau$ and the coordinate time $t$. Then, the time lapse is obtained by expanding the coordinate time as $t=\bar t + \delta t$ and taking into account that the proper-time would correspond to the time measured by the observer in a homogeneous universe, i.e., $\tau=\bar t\,$ at the exact non-perturbative level. 
In formulae,
\[
\tau (t,\textbf{x}) = t + \int_0^{\bar{t}} dt' \alpha(t',\textbf{x}) \quad
\rightarrow
\quad
\delta t =t- \bar t=t-\tau=-\int_0^{\bar{t}} dt' \alpha (t',\textbf{x})\,,
\]
and similarly for conformal time we obtain eq.~\eqref{coord time effect}.
}
Therefore, by noting the conformal time at the observer $\eta_o=\bar\eta_o + \delta \eta_o$, we have that
$
	a(\eta_{o})=a(\bar\eta_{o})[1+\mathcal H_{o}\delta\eta_{o}] ,
$
and then
\begin{equation}
\label{1+z obs}
		1+z_s = \frac{a(\bar\eta_o)}{a(\eta_s)}\big[ 1 + \mathcal H_o\delta\eta_o + \delta\Upsilon_o-\delta\Upsilon_s
		\big] \,.
\end{equation}
Furthermore, since the observed redshift $z_s$ is used to identify the time at the source in a homogeneous universe, we note the conformal time at the source as $\eta_s\equiv \bar{\eta}_z+\Delta\eta$, where the time $\bar\eta_z$ is defined as the time coordinate at the observed redshift $z_s$ and $\Delta\eta$ is the residual distortion caused by inhomogeneities. With this definition, the comoving distance to the source is
\begin{equation}
	\bar r_z \equiv \bar r (z_s) = \bar \eta_o - \bar \eta_z = \int_0^{z_s}\frac{dz}{H(z)}\,,
\end{equation}
which corresponds to the value of the affine parameter $\bar r$ at the time identified by the observed redshift $z_s$, according to the relation in eq.~\eqref{bg photon path}. Having introduced the time distortion $\Delta\eta$ at the source, we define the redshift distortion $\delta z$ by writing the observed redshift as
\begin{equation}\label{delta z}
\begin{split}
	& 1+z_s = \frac{a(\bar\eta_o)}{a(\bar\eta_z)} \equiv (1+\bar z)(1+\delta z) \,,
\qquad\qquad
1+\bar z = \frac{a(\bar \eta_o)}{a(\eta_s)}\,,
\\
\delta z = \mathcal H_o\delta\eta_o  + \delta\Upsilon_o & -\delta\Upsilon_s = \mathcal H_o\delta\eta_o + \big[\mathcal V^r - \alpha_{\chi} - H\chi \big]^s_o   
	  -  \int_0^{\bar r_z} d\bar r \, \big[\alpha_\chi - \varphi_\chi -\Psi^r - C^{rr} \big]' \,,
\end{split}
\end{equation}
where $1+\bar z$ corresponds to the background expansion, while the redshift distortion $\delta z$ (related to the time distortion $\Delta\eta$) represents the effect of inhomogeneities. Given a coordinate transformation, the scale factor is related as
\begin{equation}
	\hat \eta_s = \eta_s + T_s \,, \quad\qquad a(\hat \eta_s) = a(\eta_s)\,[1+\mathcal H _s T_s] \,, \quad\qquad 1+\hat{\bar z}= (1+\bar z)\,(1-\mathcal H _s T_s)\,,
\end{equation}
and the gauge transformation of the redshift distortion is
\begin{equation}
	\delta\hat z=\delta z + \mathcal H_s T_s \,.
\end{equation}
Naturally, the reciprocal cancellation of these gauge modes proves the gauge-invariance of the observed redshift derived with the GLC approach.

At this point, by expanding the scale factor at the source as $a(\eta_s)=a(\bar{\eta}_z)[1+\mathcal H_s\Delta\eta]$, from eq.~\eqref{1+z obs} we obtain 
\begin{equation}
\label{gi redshift}
1+z_s = \frac{a(\bar\eta_o)}{a(\bar\eta_z)}\big[ 1 + \delta z - \mathcal H_s\Delta\eta \big]\,, \qquad\qquad\qquad \delta z = \mathcal H_s \Delta \eta \,,
\end{equation}
which yields the relation between time and redshift distortions.

We noticed that in the previous works on the GLC approach and its applications, the coordinate time lapse $\delta\eta_o$ was neglected. The primary aim of those works was to obtain the second order fluctuation in the luminosity distance, where terms at the observer are not dominant. However, as we showed above, the time lapse at the observer is essential for ensuring the gauge-invariance of the observed redshift and, as we shall see, of all light-cone observables.

\subsection{Luminosity distance}
\label{Luminosity Distance}

As already mentioned, the observed position and the redshift of source galaxies
are affected by the matter fluctuations and the gravitational waves
between the source galaxies and the observer. The observed
flux of the source galaxies is also affected by the same fluctuations
and this effect is described by the fluctuation $\delta\mathcal D_L$ in the
luminosity distance $\mathcal D_L=\bar{\mathcal D}_L (1+\delta \mathcal D_L)$, where $\bar{\mathcal D}_L(z_s)=(1+z_s)\,\bar r_z$ is the luminosity distance in a homogeneous universe. 

Let us recall that the luminosity distance $\mathcal D_L$ of a source at redshift $z_s$ is related to the angular diameter distance $\mathcal D_A$ by:
\begin{equation}\label{relation dA dL}
\mathcal D_L=(1+z_s)^2\,\mathcal D_A \,.
\end{equation}
With this exact relation, the perturbation of the angular diameter distance and of the luminosity distance are identical. Therefore, the fluctuation in the luminosity distance can be obtained by computing the angular diameter distance. 
In the unperturbed background and for a source at redshift $z_s$ the angular diameter distance is simply given by $\bar{\mathcal D}_A (z_s)  = a(\bar\eta_z)\,\bar r_z\,$.
When taking inhomogeneities into account this well known result is modified and the angular diameter distance can be obtained by considering a physical area $d A$ perpendicular to the light propagation in the rest frame of the source. This infinitesimal area would appear subtended by a solid angle $d\Omega_{\text{obs}}= \sin\theta_{\text{obs}} \, d\theta_{\text{obs}} \, d\phi_{\text{obs}}$ measured by the observer in the rest frame, and it is related to the angular diameter distance as 
$
dA=\mathcal{D}_A^2 \, d\Omega_{\text{obs}}  \,.
$   

In GLC coordinates, the area perpendicular to the photon wavevector at the source position is given by
\begin{equation}
\label{dA}
	dA=\mathcal D_A^2 \, d\Omega_{\text{obs}} = \sqrt{\vert \gamma \vert}\,d^2 \tilde\theta \,.
\end{equation}
This quantity also represents a measure on the two-sphere identified by the redshift $z_s$ and parametrized by $\tilde\theta^a$, where $\gamma_{ab}$ is the induced metric. Such measure can be used to average scalar quantities on the constant redshift two-sphere embedded in the observer past light-cone, according to the prescription introduced in \cite{GLC2011}:
\begin{equation}
\label{average}
\langle S \, \rangle_{w,z_s} \equiv \frac{\int d^2\tilde\theta \, \sqrt{\vert \gamma(w,\tau_s,\tilde\theta^a)\vert}\, S(w,\tau_s,\tilde\theta^a)}{\int d^2\tilde\theta \, \sqrt{\vert \gamma (w,\tau_s,\tilde\theta^a)\vert}} = \frac{\int dA \, S}{\int dA}\,,
\end{equation}
where $S$ is a generic scalar.
From eq.~\eqref{dA}, the measure $d^2\tilde\theta \sqrt{\vert \gamma \vert}$ is expressed in terms of the angular diameter distance and the observed solid angle (both gauge-invariant quantities) yielding a gauge-invariant prescription for the light-cone average.
We also note that the physical area element in GLC coordinates ($dA=\sqrt{\vert \gamma \vert}\,d^2 \tilde\theta\,$) does not depend on how we fix the degrees of freedom in the GLC angles (see sec.~\eqref{sec Gauge-invariance of the GLC observables} below eq.~\eqref{theta}). Indeed, when no condition is imposed, the GLC angles are generally given by the observed angles plus a constant at the observer. As a consequence, the differentiation of the GLC angles is the same whatever value the constant at the observer has, leaving the physical area unaffected by our choice for the GLC angles.
Regarding the angular diameter distance, as we show in appendix \ref{Identification of the GLC angles with the observed angles}, when the GLC angles are matched to the observed angles, $\tilde\theta^a=(\theta_{\text{obs}},\phi_{\text{obs}})$, eq.~\eqref{dA} reduces to the simple formula
\begin{equation}
\label{DA formula}
\mathcal D_A^2 = \frac{\sqrt{\vert\gamma\vert}}{\sin\tilde\theta}  \,.
\end{equation}
On the other hand, when no condition is imposed to fix the degrees of freedom in $\tilde\theta^a$, the angular diameter distance is generally given by 
\begin{equation}
\label{ang dist}
	\mathcal D_A^2 = \sqrt{\vert \gamma \vert} \, \frac{d^2 \tilde\theta}{d\Omega_{\text{obs}}} \,.
\end{equation}
We are now going to calculate the expression of $\mathcal D_A$, demonstrating that indeed the final result does not depend on our choice of angles.
From eq.~\eqref{GAMMA}, the determinant $\gamma=\text{det}\,\gamma_{ab}$ is given by
\begin{equation}
\label{det gamma}
	\begin{split}
		\gamma &= a^4 r^4  \sin^2\theta \,\big[ 1 + 4\,(\varphi_\chi + H\chi) -2\,\partial_a\,\delta\tilde\theta^a + 2\, \bar g_{ab} (\mathcal G^{a|b}+C^{ab})
		\big] \,.
\end{split}
\end{equation}
Note that to the first order in perturbations the determinant is $\gamma=\gamma_{11}\gamma_{22}$, because the off-diagonal entries contain only first order terms and their product would be of second order. Furthermore, for these diagonal matrix elements the operator $\partial_a$ commutes with $\bar g^{ab}$.
After substituting the expression of $\gamma$ in eq.~\eqref{det gamma}, we can write the angular diameter distance as
\begin{equation}
\begin{split}
	\mathcal D^2_A =  a_s^2 r_s^2  \big[ 1+2\,(\varphi_\chi + H\chi)-\partial_a \delta\tilde\theta^a + \bar g_{ab}(\mathcal G^{a|b}+C^{ab}) \big]  \frac{\sin\theta_s}{\sin\theta_{\text{obs}}} \frac{d \tilde\theta d \tilde\phi}{d\theta_{\text{obs}}\,d\phi_{\text{obs}}}\,.
\end{split}
\end{equation}
The last factor (which is unity if the GLC angles are matched to the observed angles) can be conveniently written as
\begin{equation}\label{a}
\begin{split}
\frac{d \tilde\theta d \tilde\phi}{d\theta_{\text{obs}}d\phi_{\text{obs}}} = \frac{d \tilde\theta d \tilde\phi}{d\theta d\phi} \times \frac{d \theta d\phi}{d\theta_{\text{obs}}\,d\phi_{\text{obs}}}\,,
\end{split}	
\end{equation}
and the two Jacobian determinants of the transformations $\theta^a \rightarrow \tilde\theta^a$ and $\theta^a_{\text{obs}} \rightarrow \theta^a$ can be calculated according to the relations between the different angles ($\theta^a = \theta^a_{\text{obs}}+\delta\theta^a$,\, $\tilde\theta^a = \theta^a +\delta\tilde\theta^a$):
\begin{equation}\label{c}
\begin{split}
	\frac{d \tilde\theta d \tilde\phi}{d\theta d\phi} &= \det\bigg[\frac{\partial \tilde\theta^a}{\partial\theta^b}\bigg]= \det\bigg[\frac{\partial(\theta^a+\delta\tilde\theta^a)}{\partial\theta^b}\bigg]=1+\partial_a \delta\tilde\theta^a \,,
	\\
	\frac{d \theta d\phi}{d\theta_{\text{obs}}\,d\phi_{\text{obs}}} &= \det\bigg[\frac{\partial \theta^a}{\partial\theta_{\text{obs}}^b}\bigg]=\det\bigg[ \frac{\partial (\theta^a_{\text{obs}}+\delta\theta^a)}{\partial\theta^b_{\text{obs}}}\bigg] = 1 + \frac{\partial}{\partial\theta_{\text{obs}}}\delta\theta + \frac{\partial}{\partial\phi_{\text{obs}}}\delta\phi \,.
\end{split}
\end{equation}
Therefore, the angular diameter distance becomes
\begin{equation}
\label{DA theta obs}
	\mathcal D^2_A =  a_s^2 r_s^2  \big[ 1+2\,(\varphi_\chi + H\chi) + \bar g_{ab}(\mathcal G^{a|b}+C^{ab}) \big] \, \frac{\sin(\theta_{\text{obs}}+\delta\theta)}{\sin\theta_{\text{obs}}} \, \bigg[ 1 + \frac{\partial}{\partial\theta_{\text{obs}}}\delta\theta + \frac{\partial}{\partial\phi_{\text{obs}}}\delta\phi \bigg]\,,	
\end{equation}
where the last two factors are related to the gravitational lensing convergence $\kappa$ as
\begin{equation}
1-2\,\kappa = \frac{\sin(\theta_{\text{obs}}+\delta\theta)}{\sin\theta_{\text{obs}}} \bigg[ 1 + \frac{\partial}{\partial\theta_{\text{obs}}}\delta\theta + \frac{\partial}{\partial\phi_{\text{obs}}}\delta\phi \bigg]  \,.
\end{equation}
Since the above expression does not contain GLC variables, it cannot be calculated within the GLC approach here.\footnote{In \cite{Fanizza:2014baa} the GLC metric was employed to derive exact and non-perturbative expressions of lensing quantities such as shear and optical scalars.} Instead, we can use the geometric approach described in \cite{Yoo}, which gives
\begin{equation}
\begin{split}
\kappa  =&  [-\mathcal V^r + \Psi^r + C^{rr}]_o  + \frac{1}{2}\hat\nabla_a  \mathcal G^a  + \frac{1}{\bar r_z}\mathcal G^r_o
\\
&+ \frac{1}{2}  \int_0^{\bar r_z}d\bar r \bigg[ \hat\nabla_a \big(\Psi^a +2\,C^{ra}\big) + \bigg(\frac{\bar r_z-\bar r}{\bar r_z \bar r}\bigg) \hat\nabla^2 (\alpha_\chi - \varphi_\chi - \Psi^r - C^{rr}) \bigg]  \,,
\end{split}
\end{equation}
where $\hat\nabla_a  \mathcal G^a = \partial_a \mathcal G^a + \cot\theta\,\mathcal G^\theta$. The same result is derived in appendix \ref{appendix B}, where the GLC angles are matched to the observed angles. This quantity, describing the convergence of light rays due to the effect of inhomogeneities between source and observer, gauge transforms as 
\begin{equation}
	\hat \kappa = \kappa  -  \frac{1}{2}\hat\nabla_a  \mathcal L^a_s - \frac{1}{\bar r_z}\mathcal L^r_o \,.
\end{equation}
Then, after taking the square root of eq.~\eqref{DA theta obs} root we have
\begin{equation}
	\mathcal D_A(\lambda_s) =  a_s r_s  \big[ 1 -\kappa + \Xi \, \big]  \,, \qquad \qquad \Xi = \frac{1}{2} (\mathcal C^{\alpha}_\alpha - \mathcal C_{\alpha\beta} n^\alpha n^\beta) \,,	
\end{equation}
where $ n^\alpha=(1,0,0)$ is a unit directional vector representing the light propagation direction in a homogeneous universe.
At this point, to complete our derivation, we only need the expression for $a_s r_s$ to first order. As in \cite{GLC2012}, by applying eq.~\eqref{omega GI} to the observer light-cone $w=\eta_o$ evaluated at the source position, we get
\begin{equation}
\label{omega 0}
	w_s = \eta_s+r_s - \bar r_z \Psi_{\text{av}} = \eta_o \,,
\end{equation}
where we have denoted the average of the perturbations along the unperturbed null geodesic as
\begin{equation}
\begin{split}
\Psi_{\text{av}} &\equiv \frac{1}{\bar r_z}\int_0^{\bar r_z} d\bar r\, [\mathcal A - \mathcal B^r - \mathcal C^{rr}]  = \frac{1}{\bar r_z}\int_0^{\bar r_z} d\bar r\,[\alpha_\chi-\varphi_\chi -\Psi^r- C^{rr}]-\frac{1}{\bar r_z} \bigg[\frac{\chi}{a}+\mathcal G^r \bigg]^{s}_{o} \,.
\end{split}
\end{equation}
Now from eqs.~\eqref{gi redshift} and \eqref{omega 0} we can determine the radial coordinate $r_s$ of the source:
\begin{equation}
	r_s = \bar\eta_o - \bar\eta_z  + \delta\eta_o -\frac{\delta z}{\mathcal H_s} + \bar r_z \Psi_{\text{av}} = \bar r_z\bigg[1 + \frac{\delta \eta_o}{\bar r_z}  -\frac{\delta z}{\mathcal{H}_s \bar r_z} +  \Psi_{\text{av}} \bigg] \equiv \bar r_z + \delta r \,.
\end{equation}
As a result, we can identify the perturbation $\delta r$ of the radial coordinate (see also \cite{Yoo,Yoo:2009,Yoo:2010,Jeong:2011as}):
\begin{equation}
\begin{split}
\frac{\delta r}{\bar r_z} &= \frac{\delta \eta_o}{\bar r_z} -\frac{\delta z}{\mathcal{H}_s\bar r_z}  +   \Psi_{\text{av}} \,,
\end{split}
\end{equation}
whose gauge transformation is $\delta \hat r = \delta r + \mathcal L^r \vert^s_o\,$.
Similarly we can obtain $a_s$, indeed from eq.~\eqref{gi redshift} we have
\begin{equation}
a_s=a(\bar\eta_z)+\Delta\eta\, a'(\bar\eta_z)=a(\bar\eta_z)[1+\mathcal{H}_z\Delta\eta]=a(\bar\eta_z)[1+\delta z]\,.
\end{equation}
Therefore, we finally get the expression of $a_sr_s$ on the 2-sphere identified by $z_s$:
\begin{equation}
\begin{split}
\label{asrs}
a_sr_s &=  a(\bar\eta_z)\bar r_z \bigg[1 + \delta z + \frac{\delta r}{\bar r_z} \bigg] \,.
\end{split}
\end{equation}
Going back to the angular diameter distance we obtain
\begin{equation}
\label{da sec 3.2}
\begin{split}
\mathcal D_A &= \bar{\mathcal D}_A  \, \bigg[\, 1 + \delta z +\frac{\delta r}{\bar r_z}  -\kappa + \Xi \, \bigg] \,,
\end{split}
\end{equation}
and finally, from eq.~\eqref{relation dA dL},
\begin{equation}
\label{dL fluct}
\delta \mathcal D_A = \delta\mathcal D_L = \delta z +\frac{\delta r}{\bar r_z}  -\kappa + \Xi   \,.
\end{equation}
This covariant expression is fully consistent with the luminosity distance fluctuation derived in \cite{Yoo} with the geometric approach and in a general metric representation. This result also perfectly matches the luminosity distance calculated with other approaches but with specific choice of gauge conditions (see \cite{Yoo:2016}). By taking the gauge transformation of the various terms we obtain
\begin{equation}
	\begin{split}
	\delta \hat{\mathcal D}_A = \delta \hat{\mathcal D}_L &= \delta \hat z + \frac{\delta \hat r}{\bar r_z} - \hat\kappa + \frac{1}{2}(\hat{\mathcal C}^{\alpha}_{\alpha}-\hat{\mathcal C}_{\alpha \beta}n^\alpha n^\beta)
	\\
	&=  (\delta z + \mathcal H_s T_s) + \bigg(\frac{\delta r}{\bar r_z} + \frac{1}{\bar r_z}\mathcal L^r \big\vert^s_o \bigg) -\bigg(\kappa  -  \frac{1}{2}\hat\nabla_a  \mathcal L^a_s - \frac{1}{\bar r_z}\mathcal L^r_o \bigg)
	\\
	&\quad + \bigg( \frac{1}{2} (\mathcal C^{\alpha}_{\alpha}-\mathcal C^{\alpha}_{\beta}n^\beta)  -  \mathcal H_s T_s - \frac{1}{\bar r_z}\mathcal L^r_s -  \frac{1}{2}\hat\nabla_a  \mathcal L^a_s \bigg) = \delta \mathcal D_A    = \delta \mathcal D_L \,.
	\end{split}
\end{equation}
The cancellation of gauge modes among different terms is shown explicitly, demonstrating the gauge-invariance of the angular diameter distance and the luminosity distance in the GLC approach. 

The above derivation shows that the expression of the luminosity distance is independent of the normalization of the GLC angles at the observer position. Indeed, the Jacobian of the transformation from the GLC angles to the observed angles cancels the terms related to the GLC angular distortions $\delta\tilde\theta^a$. In this way, the nature of the GLC angles becomes irrelevant for the derivation of the luminosity distance. 
To demonstrate this statement, we derive in appendix \ref{Identification of the GLC angles with the observed angles} the angular diameter distance after fixing the degrees of freedom in the GLC angles to match the observed angles (measured in the observer rest frame). In this case the angular diameter distance is simply given by eq.~\eqref{DA formula} and the calculation of the gravitational lensing convergence can be performed in the GLC approach, as described in appendix \ref{appendix B}.

\subsection{Physical volume}
\label{Galaxy Number Density}
  
Due to the presence of inhomogeneities the volume $V_{\text{obs}}$ inferred from the observed redshift and angle does not correspond to the \textit{physical} volume $V$  occupied by the source galaxies. 
To account for this effect, we define the volume distortion $\delta V$, such that $dV = (1+\delta V)\,dV_{\text{obs}}$. The volume distortion is a gauge-invariant quantity, as we demonstrate in this section after deriving its expression with the GLC approach.
 
In \cite{Yoo} the infinitesimal physical volume occupied by the source is written in terms of the observed redshift $z_s$ and angles $\theta_{\text{obs}}$, $\phi_{\text{obs}}$ :
\begin{equation}
\label{dV}
	d V = \sqrt{-g}\,\epsilon_{\mu\nu\rho\sigma}\, u^\mu_s \, dx^\nu dx^\rho dx^\sigma = \sqrt{-g}\,\epsilon_{\mu\nu\rho\sigma}\, u^\mu_s \, \frac{\partial x^\nu}{\partial z_s} \frac{\partial x^\rho}{\partial \theta_{\text{obs}}} \frac{\partial x^\sigma}{\partial \phi_{\text{obs}}}dz_s d\theta_{\text{obs}} d\phi_{\text{obs}} \,.
\end{equation}
On the other hand, the inferred volume is given by
\begin{equation}
\label{dVobs}
	dV_{\text{obs}}= a(\bar\eta_z)^3 \, \bar r_z^2 \, d\bar r_z \, d\Omega_{\text{obs}} = \frac{\bar r_z^2\, dz_s \, d\Omega_{\text{obs}}}{H_s (1+z_s)^3}\,,
\end{equation}
where we set $a(\bar \eta_o)\equiv 1$, so that $a(\bar \eta_z)=1/(1+z_s)$.

In GLC coordinates, the physical volume element occupied by the source is simply given by
\begin{equation}
\label{dV GLC}
	dV = dA\,d\tau= \sqrt{\vert \gamma \vert}\, d^2\tilde \theta \, d\tau \,.
\end{equation} 
To compare our result with that found in \cite{Yoo}, we can change the GLC coordinates into the observed variables $\theta_{\text{obs}}$,$\phi_{\text{obs}}$ and $z_s$. 
As explained in sec.~\ref{Luminosity Distance}, the differentiation of the GLC angles already corresponds to the differentiation of the observed angles ($d^2\tilde\theta = d\theta_{\text{obs}}\, d\phi_{\text{obs}}$), therefore, we only need to change variable from the proper-time $\tau$ to the observed redshift $z_s$, obtaining
\begin{equation}
	dV = -\sqrt{\vert \gamma \vert}\,\frac{\partial\tau}{\partial z_s} \, d^2\tilde \theta \, dz_s \,,
\end{equation}
where the minus sign is due to the fact that when the proper-time increases the redshift decreases and vice versa.
Let us now derive the volume distortion by calculating the physical volume element. After substituting the expression of $\gamma$ in eq.~\eqref{det gamma} and the expansion of the factor $a_sr_s$ in eq.~\eqref{asrs} we obtain 
\begin{equation}
\begin{split}
	dV &=  - a(\bar\eta_z)^2 \bar r_z^2 \,\bigg[1+2\,\delta z + 2 \, \frac{\delta r}{\bar r_z} - 2\,\kappa + 2\,\Xi\, \bigg] \frac{\partial\tau}{\partial z_s}  \, dz_s \, d\Omega_{\text{obs}}
	\\
	&= -\bigg[1+2\,\delta z + 2 \, \frac{\delta r}{\bar r_z} - 2\,\kappa + 2\,\Xi \, \bigg] \frac{\partial\tau}{\partial z_s} \, \frac{\bar r_z^2 \, dz_s \, d\Omega_{\text{obs}}}{(1+z_s)^2}\,.
\end{split}
\end{equation}
At this point what we need to compute is the change of the proper-time with respect to the observed redshift, $\partial\tau/\partial z_s$.  To simplify the calculation we rewrite this derivative as  
\begin{equation}
	\frac{\partial\tau}{\partial z_s} = \frac{\partial \tau}{\partial \eta_s} \frac{\partial \eta_s}{\partial z_s} = - \frac{\partial \tau}{\partial \eta_s}\frac{1}{H_s} \,.
\end{equation}
After expanding the emission time as $\eta_s=\bar\eta_z + \Delta\eta$, we can express the proper-time at emission as
\begin{equation}
	\tau = a(\bar\eta_z)\,\frac{\delta z}{\mathcal H_s} + \int_{0}^{\bar\eta_z} d \eta \, a(\eta)[1+\alpha] \,,
\end{equation}
obtaining 
\begin{equation}
	\frac{\partial \tau}{\partial \eta_s} = a(\bar\eta_z) \bigg[1+\alpha + \delta z - \frac{\mathcal H_s'}{\mathcal H_s^2}\,\delta z + \frac{1}{\mathcal H_s}\delta z'  \bigg] \,.
\end{equation}
Therefore, going back to the volume element, we have
\begin{equation}
\begin{split}
	dV &=  \bigg[1+3\,\delta z + \mathcal A +  \mathcal C^\alpha_\alpha + 2\, \frac{\delta r}{\bar r_z} - 2\,\kappa  - \frac{\mathcal H_s'}{\mathcal H_s^2}\,\delta z + \frac{1}{\mathcal H_s}\delta z'  - \mathcal C_{\alpha\beta} n^\alpha  n^\beta \bigg] \, \frac{\bar r_z^2 \, dz_s \, d\Omega_{\text{obs}}}{H_s(1+z_s)^3}\,.
\end{split}
\end{equation}
The above equation can be further simplified by noting that
\begin{equation}
	- \frac{\mathcal H_s'}{\mathcal H_s^2}\delta z + \frac{1}{\mathcal H_s}\delta z' = H_s \frac{\partial}{\partial z_s} \delta r  + V_\alpha  n^\alpha -  \mathcal A + \mathcal C_{\alpha\beta} n^\alpha  n^\beta \,.
\end{equation}
In this way the volume element becomes
\begin{equation}
dV =  \bigg[1+3\,\delta z  +  \mathcal C^\alpha_\alpha + 2\, \frac{\delta r}{\bar r_z} - 2\,\kappa +  H_s \frac{\partial}{\partial z_s} \delta r + V_\alpha  n^\alpha \bigg] \,  dV_{\text{obs}}	\,.
\end{equation}
As a result, the final expression for the volume distortion is
\begin{equation}
\label{delta V}
	\delta V = 3\,\delta z  +  \mathcal C^\alpha_\alpha + 2\, \frac{\delta r}{\bar r_z} - 2\,\kappa +  H_s \frac{\partial}{\partial z_s} \delta r + V_\alpha  n^\alpha \,.
\end{equation}
This quantity is covariant and gauge-invariant, besides it coincides with the result found in \cite{Yoo}. If compared with the volume distortion derived in \cite{DiDio} with the GLC approach and in the conformal Newtonian gauge, this result includes perturbations at the observer not considered there, but crucial for the gauge-invariance of the final expression.

\section{Discussion}
\label{Conclusions}

In this work we showed \textit{explicitly} the gauge-invariance of light-cone observables derived in the GLC approach. We also considered the full general metric to first order in perturbations for the first time within the GLC formalism. Furthermore, by comparing the results with those derived in the approach introduced in \cite{Yoo:2009,Yoo:2010,Yoo}, we demonstrated the full consistency of the two methods to calculate expressions of light-cone observables in the presence of inhomogeneities in the Universe. Our study provides further understanding of the properties of the GLC representation.

First of all, in sec.~\ref{sec Gauge-invariance of the GLC observables} we pointed out the presence of new degrees of freedom in the expression of the GLC angles, given by perturbations evaluated at the observer position. These angular degrees of freedom are also studied in \cite{Fleury:2016}, with a discussion about how they can be fixed to describe different physical situations. As we show in appendix \ref{Identification of the GLC angles with the observed angles}, by fixing the degrees of freedom through a proper normalization, the GLC angles can be identified with the observed angles, measured by the observer in the rest frame. On the other hand, a different normalization at the observer position is possible, leading to a different form of the GLC angles, which would then correspond to the observed angles and a constant at the observer. Naturally, the final expressions of light-cone observables cannot depend on our choice of normalization. To demonstrate this point, in sec.~\ref{Luminosity Distance} we derived the gauge-invariant expression of the luminosity distance without fixing the degrees of freedom in the GLC angles. The same result is obtained in appendix \ref{Identification of the GLC angles with the observed angles}, where a specific normalization is taken instead. Such normalization, according to which the GLC angles match the angles measured by the observer in the rest frame, is probably the most convenient, as it leads to a very simple formula for the angular diameter distance, eq.~\eqref{DA formula}. When a different normalization is chosen, the formula of the angular diameter distance contains an additional factor given by the Jacobian of the rotation from the GLC to the observed angles. However, when the GLC angles appear under differentiation, as in the physical area and volume occupied by the source, the difference becomes completely irrelevant since the differentiation of any constant at the observer (representing the difference between GLC angles and observed angles) would vanish.

In \cite{GLC2012,BenDayan:2012,BenDayan:2013}, the luminosity distance in the presence of inhomogeneities is derived from the angular diameter distance in eq.~\eqref{DA formula}. However, the difference between the observed angle in a GLC coordinate and that in the observer rest frame was not considered, as well as the presence of degrees of freedom in GLC angular coordinate at the observer. 
  If the difference between the angle in the observer rest frame and that in a global coordinate is neglected, the degrees of freedom are automatically set to zero and the GLC angular coordinate does not match the angle in the observer rest frame. This results in the absence of some terms in the final expression for the luminosity distance, such as the observer peculiar velocity and the gravitational potential at the observer position. Without these terms the luminosity distance is not gauge-invariant and not consistent with the equivalence principle (see \cite{Biern:2016kys}). In \cite{GLC2013} the normalization condition for the angular GLC variables was fixed in the expression of the angular diameter distance by a factor evaluated at the observer, which can be interpreted as the Jacobian of the rotation from a generic GLC angular coordinate to the observed angle in the observer rest frame. 
   
In sec.~\ref{Redshift parameter} we derived the observed redshift, stressing the importance of including the time lapse at the observer. This term represents the effect due to the fact that the observer proper-time does not correspond to the coordinate time in the physical universe. Indeed, the presence of inhomogeneities induces a perturbation in the coordinate time at observation, which is captured by the time lapse. Specifically, the inhomogeneities affect the observer four-velocity, causing a discrepancy between the time measured and the coordinate time. As we showed in sec.~\ref{Redshift parameter}, only if the time lapse at the observer is included the expression of the redshift is gauge-invariant. This argument is later extended to any light-cone observable, as the time lapse appears not only in the redshift distortion but also in the distortion of the radial distance between source and observer.

In sec.~\ref{Luminosity Distance}, in order to obtain the angular diameter distance, we made use of the fact that the infinitesimal area $dA$ occupied by the source is equal to the measure $\sqrt{\gamma}\,d^2\tilde\theta$ on the fixed-time two-sphere embedded in the light-cone. This equality results directly in the gauge-invariance of the light-cone average prescription introduced in \cite{GLC2011}. Given the gauge-invariance of the light-cone average, this can be applied to compute the mean of observables in the presence of inhomogeneities, as it has been done in \cite{BenDayan:2013,BenDayan:2012ct} (and partially in \cite{GLC2012}). 
Indeed, deriving the full relativistic expression of a given observable is not enough to interpret the outcome of a survey. Consider for instance the relation between the luminosity distance $\mathcal D_L$ and the observed redshift $z_s$ of a given source. 
 As described in \cite{Fleury:2016fda}, the observational strategy consists in collecting many data points $(z_s,\mathcal D_L)$, and the value of $\mathcal D_L$ at a given redshift $z_s$ is obtained by averaging over the data in the redshift bin containing $z_s$. Consequently, also the theoretical expression of the luminosity distance as a function of the observed redshift needs to be averaged. To this purpose, second-order calculations are needed (see \cite{BenDayan:2013,BenDayan:2012ct,Ben-Dayan:2014swa}). The study of the GLC formalism in this work can also be used to go beyond the linear order, providing the correct starting point for the derivation and a concrete way to use the observed angles in the GLC angular coordinate, being this the most physically meaningful choice. 
 
Finally, in sec.~\ref{Galaxy Number Density} we derived the expression of physical volume occupied by sources, obtaining the volume distortion due to relativistic effects. The importance of a precise theoretical derivation of the volume distortion relies on the fact that this latter is used to predict the number density of galaxies, which is a key observable to test different cosmological models. The observed galaxy number density is obtained by counting the number of galaxies in the observed redshift range and within the observed solid angle.
Whereas the observed volume occupied by the source galaxies is different from the physical volume, the number of galaxies within the volume is not affected by the inhomogeneities. As a consequence, by calculating the volume distortion we can relate the observed galaxy number density to the predicted physical one.

In summary, the GLC approach, if exercised properly, results in the correct and consistent expressions of light-cone observables. It also offers a covariant and gauge-invariant prescription for averaging scalars on our past light-cone, providing a simple way to estimate the effect of inhomogeneities on the observables that are measured in large scale structure surveys.

\acknowledgments

We thank Sang Gyu Biern, Giuseppe Fanizza and Ermis Mitsou for useful discussions. In addition, we would like to express our gratitude to Gabriele Veneziano, Giovanni Marozzi and Maurizio Gasperini for providing beneficial comments about this work. We acknowledge support by the Swiss National Science Foundation, and J.Y. is further supported by a Consolidator Grant of the European Research Council (ERC-2015-CoG grant 680886). 

\appendix

\section{Technical details}
\label{appendix A}

In this short appendix, we provide the covariant derivatives of the metric perturbations and useful relations to simplify our calculations in the main text. 

First of all, given the background 3-spatial metric tensor $\bar g_{\alpha\beta}$ in spherical coordinates, the affine connections are readily derived as
\begin{equation}
	\begin{split}
		&\Gamma^r_{rr}=\Gamma^r_{ra}=0 \,, \quad\qquad \Gamma^r_{ab}=-\frac{1}{r}\,\bar g_{ab} \,, \qquad \Gamma^a_{rr}=0 \,, \quad\qquad \Gamma^a_{rb}=\frac{1}{r}\,\delta^a_b \,,
		\\
		&\Gamma^\theta_{\theta\theta}=\Gamma^\theta_{\theta\phi}=\Gamma^\phi_{\theta\theta}=\Gamma^\phi_{\phi\phi}=0\,, \quad\quad \Gamma^\theta_{\phi\phi}=-\sin\theta\cos\theta \,,\quad\quad \Gamma^\phi_{\theta\phi}=\cot\theta \,,	
	\end{split}
\end{equation}
where $\delta^a_b$ is the Kronecker delta. As a result, the covariant derivatives can be expressed in terms of ordinary derivatives as
\begin{equation}
	\begin{split}
	 &\gamma^{,r|r}=\gamma^{,rr}\,, \qquad\qquad \gamma^{,r|a}=\gamma^{,a|r}=\gamma^{,ra}-\frac{\gamma^{,a}}{r}  \,,
	\\	
	&C^{r|r}=C^{r,r}\,, 	\qquad\qquad C^{r|a}=C^{r,a}-\frac{C^a}{r}\,, \qquad\qquad C^{a|r}=C^{a,r}+\frac{C^a}{r}\,, \quad\quad 
	\\
	&\bar g_{ab} (\gamma^{,a|b}+C^{a|b}) = \partial_a[\gamma^{,a}+C^a] +\cot\theta \,[\gamma^{,\theta}+C^\theta]+\frac{2}{r}[\gamma^{,r}+C^r] \,.
\end{split}
\end{equation}
It is important to note the distinction
\begin{equation}
	\begin{split}
	\gamma^{,ar}=\gamma'^{,a} +\frac{d}{d\bar r}\gamma^{,a} \,,
	\qquad\qquad
	\gamma^{,ra}=\gamma'^{,a}+\frac{2}{r}\gamma^{,a} +\frac{d}{d\bar r}\gamma^{,a}\,.
	\end{split}
\end{equation}
Indeed, the derivatives $\partial^r$ and $\partial^a$ do not commute and therefore $\gamma^{,ra}\neq \gamma^{,ar}$, instead $[\partial^a,\partial^r]\gamma=2\gamma^{,a}/r$. 
Second, in the calculations performed throughout the paper we used the following formulas for double integrations:
\begin{equation}
\begin{split}
&\int_0^{\bar r_z} d\bar r \int_0^{\bar r}d\bar{r}'\, f(\bar{r}') = \int_0^{\bar r_z}d\bar r \, (\bar r_z-\bar r)f(\bar r)  \,,
\\	
&\int_0^{\bar r_z} d\bar r\, \frac{1}{\bar r\,^2} \int_0^{\bar r}d\bar{r}'\, f(\bar{ r}') = \int_0^{\bar r_z}d\bar r \, \bigg(\frac{\bar r_z - \bar r}{\bar r_z \bar r}\bigg)f(\bar r)+ f(0) \,,
\end{split}
\end{equation}
where $f(x)$ is a generic function of $x$.

\section{Matching conditions for the GLC angles} 
\label{Identification of the GLC angles with the observed angles}

In this appendix we show how to fix the degrees of freedom in the GLC angles to match them with the observed angles (in the observer rest frame). Then, we will derive the angular diameter distance under this condition, showing that we obtain the same result of sec.~\eqref{Luminosity Distance}. 

The degrees of freedom which we have at hand are associated with the quantities $\delta\tilde\theta^a_o$ and $ \delta w_o^{,a}$ in the expression of the GLC angles $\tilde\theta^a$, eq.~\eqref{theta}. Using the exact relation $k^\mu=g^{\mu\nu} \partial_\nu w\,$ we relate $ \delta w_o^{,a}$ to the wavevector perturbation $\delta k^a_o$ as
\begin{equation}
	\bar g^{ac} \partial_c \, \delta w_o =  [a^2\delta k^a]_o + \mathcal B^a_o + 2\,\mathcal C^{ra}_o \,.
\end{equation}
In this case the GLC angles become
\begin{equation}
\label{theta when GLC=obs}
\begin{split}
\tilde\theta^a_s  &=  \theta^a_s +\delta\tilde\theta^a_o - \bar r_z \bigg[a^2 \delta k^a + \frac{d}{d\bar r}\mathcal G^a + \Psi^a+2\,C^{ra} \bigg]_o + \mathcal G^a \big\vert^s_o  
\\
 &\quad + \int_0^{\bar r_z} d\bar r \bigg[ \Psi^a+2\,C^{ra} + \bigg(\frac{\bar r_z-\bar r}{\bar r_z \bar r}\bigg) \hat{\bar{g}}^{ac}\partial_c (\alpha_\chi-\varphi_\chi-\Psi^r-C^{rr}) \bigg]
		 \,.
\end{split}
\end{equation}
Both $\delta\tilde\theta^a_o$ and $\delta k_o^a$ represent perturbations to the photon propagation direction at observation, and are the rotational degrees of freedom to set.
The observed direction of the photons, described by the observed angles $\theta^a_{\text{obs}}=(\theta_{\text{obs}},\phi_{\text{obs}})$, is  identified in the observer rest frame. Therefore, to fix $\delta\tilde\theta^a_o$ and $\delta k^a_o$ such that $\tilde\theta^a=\theta^a_{\text{obs}}$, we have to consider the photon wavevector in the observer rest frame and study how it is related to the photon wavevector in the global coordinates $y^\mu_{\text{FRW}}$, derived by coordinate transforming the GLC wavevector.
First of all, we write the GLC wavevector $k_\mu^{\text{GLC}}=(1,0,\vec{0}\,)$ in the global coordinates $y^\mu_{\text{FRW}}=(\eta,r,\theta^a)$ by taking a coordinate transformation from the GLC coordinates $x^\mu_{\text{GLC}}=(w,\tau,\tilde\theta^a)$:
\begin{equation}
 k_\mu^{\text{FRW}}  = \frac{\partial x^\nu}{\partial y^\mu}k_\nu^{\text{GLC}} = (1+\delta w' \,,\, \hat n_\alpha + \partial_\alpha \delta w) \,,
\end{equation}
\begin{equation}
	 k^\mu_{\text{FRW}}=g^{\mu\nu}_{\text{FRW}} \, k_\nu^{\text{FRW}} = \frac{1}{a^2}(-1-\delta w'+2\mathcal A - \mathcal B_\alpha \, \hat n^\alpha \,,\, \hat n^\alpha + \bar g^{\alpha\beta}\partial_\beta \delta w - \mathcal B^\alpha - 2\,\mathcal C^{\alpha}_{\beta} \, \hat n^\beta ) \,.
\end{equation}	
The unit vector $\hat n^\alpha$ is defined in the global coordinates, and identifies the photons direction in the absence of perturbations. 
By making use of the exact relation $k^\mu=g^{\mu\nu}\partial_\nu w$, we can express the wavevector in terms of the perturbations $\delta k^\mu$ which we are interested in:
\begin{equation}
\label{k from GLC}
	k^\mu_{\text{FRW}} = \frac{1}{a^2}(-1 + a^2 \delta k^0 \,,\, \hat n^\alpha + a^2 \delta k^\alpha ) \,.
\end{equation}
We want to study the relation between this result and that obtained by mapping the photon wavevector $k_{\mathrm L}^{m}=\omega_o \, (-1\,,n^i\,)$ in the observer rest frame (local Lorentz frame) into the global coordinates.
This procedure, carefully described in appendix \ref{appendix C}, involves the construction of an orthonormal basis, the tetrads $[e_m]^\mu$, connecting the observer rest frame to the global coordinates at the observer. After deriving the tetrads, the photon wavevector in the global coordinates is given by
\begin{equation}
\label{k from L}
		k^\mu_{\text{FRW}} = [e_m]^\mu k_{\mathrm L}^{m}  =  \frac{\omega_o}{a}(-1 + \mathcal A + \hat n^\alpha V_\alpha- \hat n^\alpha \mathcal B_\alpha \,,\, n^\alpha -  V^\alpha - \hat n^\beta \mathcal C^\alpha_\beta )\,,
\end{equation}
where $\omega_o$ is the observed photon frequency and $n^\alpha \sim (\theta_{\mathrm{obs}},\phi_{\mathrm{obs}})\,$ is the unit directional vector identifying the observed angular position of the source in the rest frame.
At this point, we can match the photon wavevector in eq.\eqref{k from GLC} (obtained from the GLC wavevector) evaluated at the observer position and the photon wavevector in eq.\eqref{k from L} (obtained from the rest frame wavevector). 
We are only interested in the spatial components:
\begin{equation}
 \hat n^\alpha_o + [a^2 \delta k^\alpha]_o  = (a \omega)_o \, ( n^\alpha -  V^\alpha_o - \hat n^\beta \mathcal C^\alpha_{\beta\,o} ) \,.
\end{equation}	
The quantity $(a \omega)$ is not constant in an inhomogeneous universe. Therefore, it is convenient to split it into background and perturbation part as $ a \omega = \overline{a \omega} \, (1 + \Delta \nu) $. Considering the observer position this is $ (a \omega)_o = \overline{\omega}_o \, (1 + \Delta \nu_o) $, where $a(\bar\eta_o)\equiv 1$. 
Since the real observable we deal with is the redshift of the source, which is determined by the ratio of the photon frequency at the source to the observed frequency $\omega_o$, we never need to consider the value of $\omega_o$ in practice and we can normalize its background part as $\overline{\omega}_o\equiv 1$.
In this case we have
\begin{equation}
 \hat n^\alpha_o + \delta k^\alpha_o  = (1 + \Delta \nu_o) \, n^\alpha -  V^\alpha_o - \hat n^\beta \mathcal C^{\alpha}_{\beta\,o} \,,
\end{equation}
where the unit directional vector in the global coordinates is $\hat n^\alpha_o \sim (\theta_o,\phi_o)=(\theta_{\text{obs}},\phi_{\text{obs}}) + (\delta\theta_o,\delta\phi_o)$ while that in the observer rest frame is $n^\alpha \sim (\theta_{\text{obs}},\phi_{\text{obs}})$. 
Then, the fluctuations of the photon wavevector spatial components have to be
\begin{equation}
 \delta k^\alpha _o  =  \Delta \nu_o \,n^\alpha + (n^\alpha - \hat n^\alpha_o) -  V^\alpha_o - \hat n^\beta \mathcal C^{\alpha}_{\beta\,o} \,,
\end{equation} 
where the difference in the unit directional vectors gives the angular corrections at the observer, $ ( \hat n^\alpha_o - n^\alpha ) \sim (\delta\theta_o,\delta\phi_o)$.
We can now focus on the angular components only, obtaining
\begin{equation}
\delta k^a_o  =   -\frac{1}{\bar r_z}\delta\theta^a_o -  V^a_o -  \mathcal C^{ra}_{o} \,,
\end{equation} 
regardless of the value of the constant $\Delta \nu_o$. At this point we can use the remaining degrees of freedom to compensate for the difference between the two unit directional vectors, in order to align the photons direction in the global coordinates (in a homogeneous universe) to the observed one. To do this, we simply set $\delta\tilde\theta^a_o = -\delta\theta^a_o$, and the GLC angular distortions become
\begin{equation}
\label{final delta theta}
\begin{split}
\delta\tilde\theta^a_s = & - \bar r_z \big[ - \mathcal V^a + \Psi^a + C^{[a|r]} +C^{ra} \big]_o + \mathcal G^a \big\vert^s_o 
\\
& + \int_0^{\bar r_z} d\bar r \bigg[ (\Psi^a+2\,C^{ra}) + \bigg(\frac{\bar r_z-\bar r}{\bar r_z \bar r}\bigg) \hat{\bar{g}}^{ac}\partial_c (\alpha_\chi-\varphi_\chi-\Psi^r-C^{rr}) \bigg] \,.
\end{split}
\end{equation}
This result perfectly agrees with the angular distortions $\delta\theta^a_s=(\delta\theta_s,\delta\phi_s)$ calculated in \cite{Yoo} with the geometric approach.\footnote{In \cite{Yoo} any quantity is expressed in terms of the observables measured in the observer rest frame, which are the observed redshift $z_s$ and angles $\theta^a_{\text{obs}}$.} Specifically, the GLC angular distortions $\delta\tilde\theta^a_s$ and the distortions $\delta\theta^a_s$ calculated in \cite{Yoo} are equal but with opposite sign due to definition. Indeed, in \cite{Yoo} the angular position of the source is given by $\theta^a_s=\theta^a_{\text{obs}}+\delta\theta^a_s$, where $\theta^a_{\text{obs}}$ are the observed angles and $\delta\theta^a$ are geometric distortions due to inhomogeneities. On the other hand, in the GLC approach $\tilde\theta^a_s=\theta^a_s+\delta\tilde\theta^a_s$, where the angular distortions $\delta\tilde\theta^a_s$ cancel the distortions in $\theta^a_s$ to give the observed angles, $\tilde\theta^a_s = (\theta^a_{\text{obs}}+\delta\theta^a_s) +\delta\tilde\theta^a_s = (\theta^a_{\text{obs}}+\delta\theta^a_s) - \delta\theta^a_s = \theta^a_{\text{obs}} $. 

The quantity $\delta w_o$ represents a shift in the photons' phase at the observer position due to perturbations. This constant does not affect the expressions of light-cone observables, reflecting the freedom associated with the definition of phase. 
By considering the proportionality relation between the GLC phase $w$ (coordinate transformed to FRW) and the FRW phase $\vartheta$ (constructed from that in the observer rest frame), the integration constant $\delta w_o$ is fixed. 
In a global FRW coordinate the phase is
\begin{equation}
	\vartheta = g^{\text{FRW}}_{\mu\nu}k_{\text{FRW}}^\mu x^\nu_{\text{FRW}} = (a\omega)_o \,\big\lbrace \bar\eta_o  + \bar\eta_o \,( \mathcal A - n^i V_i)_o   
	  +\delta\eta_o + \delta r_o  
	    \big\rbrace   \,,
\end{equation}
while in a GLC coordinate the phase is given by
\begin{equation}
	w = \bar \eta_o + \delta\eta_o +  \delta r_o + \delta w_o  \,,
\end{equation}
where we evaluated both phases at the observer position.
By demanding that both be proportional, i.e. $w_o = \mathds C \, \vartheta_o $, we derive the proportionality constant and the integration constant
\begin{equation}
	\mathds C = 1 / (a\omega)_o\,, \qquad\qquad\qquad   \delta w_o  = \bar\eta_o \,( \mathcal A - n^i V_i)_o  \,.
\end{equation}

To conclude this appendix, we derive the angular diameter distance in the GLC approach when the degrees of freedom in $\tilde\theta^a$ are fixed as described above, so that $\tilde\theta^a=\theta^a_{\text{obs}}$.
From the relation between the physical area occupied by the source and the angular diameter distance, $dA=\sqrt{\vert \gamma \vert}\,d^2 \tilde\theta = \mathcal D_A^2 \, d\Omega_{\text{obs}}$, this latter is given by
\begin{equation}
	\mathcal D_A^2  = \frac{\sqrt{\vert \gamma \vert}}{\sin\theta_{\text{obs}}}\, \frac{d^2 \tilde\theta}{d^2\theta_{\text{obs}}}\,.
\end{equation}
When $\tilde\theta^a=\theta^a_{\text{obs}}$, the angular diameter distance can be expressed in terms of GLC variables only, as
\begin{equation}\label{DA glc = obs}
	\mathcal D_A^2  = \frac{\sqrt{\vert \gamma \vert}}{\sin\tilde\theta}\,.
\end{equation}
After substituting $\gamma$ with the expression in eq.~\eqref{det gamma} and taking the square root, we have
\begin{equation}
\mathcal D_A =  \bar{\mathcal D}_A   \, \sqrt{\frac{\sin \theta_s}{\sin\tilde\theta}} \, \bigg[1 + \delta z + \frac{\delta r}{\bar r_z}  - \frac{1}{2}\partial_a \delta\tilde\theta^a + \Xi \, \bigg] \,,
\end{equation}
where we also used eq.~\eqref{asrs} for the expansion of the factor $a_s r_s$ in the expression of $\gamma$.
Then, by expanding the source angle as $\theta_s=\theta_{\text{obs}}+\delta\theta_s=\tilde\theta_s-\delta\tilde\theta_s$, we get
\begin{equation}
\label{last DA}
\mathcal D_A = \bar{\mathcal D}_A  \, \bigg[\,1 + \delta z + \frac{\delta r}{\bar r_z}  - J_2 + \Xi \,  \bigg] \,,
\end{equation}
where we defined the quantity
\begin{equation}
J_2 \equiv  \frac{1}{2}\partial_a \delta\tilde\theta^a + \frac{1}{2}\cot \tilde\theta \, \delta \tilde\theta =  \frac{1}{2}\hat\nabla_a  \delta\tilde\theta^a  \,.
\end{equation}
Clearly, $J_2$ (for which we followed the notation introduced in \cite{GLC2012}) corresponds to the gravitational lensing convergence $\kappa$ introduced in sec.~\ref{Luminosity Distance}. 
To compute $J_2$ we follow the approach described in appendix \ref{appendix B}, from which we obtain
\begin{equation}
\label{J2}
\begin{split}
J_2  =&  [-\mathcal V^r + \Psi^r + C^{rr}]_o  + \frac{1}{2}\hat\nabla_a  \mathcal G^a  + \frac{1}{\bar r_z}\mathcal G^r_o
\\
&+ \frac{1}{2}  \int_0^{\bar r_z}d\bar r \bigg[ \hat\nabla_a \big(\Psi^a +2\,C^{ra}\big) + \bigg(\frac{\bar r_z-\bar r}{\bar r_z \bar r}\bigg) \hat\nabla^2 (\alpha_\chi - \varphi_\chi - \Psi^r - C^{rr}) \bigg]  \,,
\end{split}
\end{equation}
where $\hat\nabla_a \Psi^a=\partial_a \Psi^a + \cot\theta \, \Psi^\theta$ and $\hat\nabla^2=[\partial_\theta^2+\cot\theta\,\partial_\theta+(\sin\theta)^{-2}\partial_\phi^2 ]$.
This result perfectly agrees with the gravitational lensing $\kappa$ obtained in \cite{Yoo}, making the result in eq.~\eqref{last DA} fully consistent with the correct expression of the angular diameter distance in eq.~\eqref{da sec 3.2}. 

In some previous works $\delta k^a_o$ and $\delta\tilde\theta^a_o$ were set to zero, corresponding to a different choice of the GLC angles. In this case the expression of the angular diameter distance in eq.~\eqref{DA glc = obs} should contain an additional factor given by the Jacobian of the rotation from the GLC angles to the observed ones (see for instance \cite{GLC2013}), providing the perturbations at observations, such as the observer peculiar velocity and the gravitational potential, which should appear in the gravitational lensing convergence.

\section{Photon wavevector from the observer rest frame to a global coordinate}
\label{appendix C}

In the observer rest frame, where the local metric is Minkowski $g^{\mathrm L}_{mn}=\eta_{mn}$, the photon wavevector is given by
\begin{equation}
\label{kL}
	k_{\mathrm L}^{m}=\omega\,(-1\,,\,n^i\,) \,, \qquad m=t,x,y,z, \quad i=x,y,z,
\end{equation}
where $\omega = \eta_{mn}u_{\mathrm L}^{m} k_{\mathrm L}^{n}$ is the photon frequency and $n^i \sim (\theta_{\mathrm{obs}},\phi_{\mathrm{obs}})$ is a unit directional vector identifying the observed angular position of the source.

To obtain the photon wavevector in a global coordinate $y^{\mu}_{\text{FRW}}$ we need to construct an orthonormal basis in the observer rest frame, the so-called tetrads $[e_m]^\mu$. First of all, the time-like observer four-velocity $u^\mu$ defines the proper-time direction in the observer rest frame 
\begin{equation}
[e_t]^\mu\equiv u^\mu \,.	
\end{equation}
Spatial hypersurfaces orthogonal to $[e_t]^\mu$ are defined by three space-like vectors $[e_i]^\mu$. To obtain the expression for the space-like tetrads $[e_i]^\mu$, we use the orthonormality condition 
\begin{equation}
\eta_{mn}=g_{\mu\nu}[e_{m}]^\mu [e_{n}]^\nu \,.	
\end{equation}
By taking the metric given in eq.~(2.5) as $g_{\mu\nu}$ and considering the spatial components of the above condition, $\delta_{ij}=[e_i]^\mu [e_j]^\nu g_{\mu\nu}$, we obtain
\begin{equation}
\label{delta ij}
	[e_i]^\alpha [e_j]^\beta (\bar g_{\alpha\beta}+2\,\mathcal C_{\alpha\beta})=\frac{1}{a^2} \delta_{ij} \,.
\end{equation}
We now make the following ansatz:
\begin{equation}
	[e_i]^\alpha \equiv \frac{1}{a}(\delta^\alpha_i + \mathcal D^\alpha_i) \,,
\end{equation}
where $\mathcal D^\alpha_i$ is a generic tensor perturbation to be determined. 
This definition (with the Kronecker delta) means that in the absence of perturbations the spatial coordinates in the rest frame are aligned to the spatial global coordinates locally at the observer position. 
By substituting the ansatz into eq.~\eqref{delta ij} we obtain that $\mathcal D_{ij} = -\mathcal C_{ij}$ and therefore
\begin{equation}
	[e_i]^\alpha = \frac{1}{a}(\delta^\alpha_i - \mathcal C^\alpha_i) \,.
\end{equation}
Finally, from the mixed time-space components of the orthonormality condition, $0=[e_t]^\mu[e_i]^\nu g_{\mu\nu}$, we obtain	
\begin{equation}
	[e_i]^\eta = \frac{1}{a}(V_i - \mathcal B_i)\,.
\end{equation}
Summing up, the tetrads are given by
\begin{equation}
	[e_t]^\mu = u^\mu \,, \qquad\qquad\qquad [e_i]^\mu = \frac{1}{a}(V_i - \mathcal B_i\,,\,\delta^\alpha_i - \mathcal C^\alpha_i)\,.
\end{equation}
As a result, the photon wavevector in a global coordinates is given by
\begin{equation}
		k^\mu_{\text{FRW}} = [e_m]^\mu k^m_L = \frac{\omega}{a}(-1 + \mathcal A + n^i V_i- n^i\mathcal B_i \,,\, \delta^\alpha_i \, n^i -  V^\alpha - n^i \mathcal C^\alpha_i )\,.
\end{equation}

It is noted that the unit directional vector $n^i$ in the observer rest frame is different from the unit directional vector $\hat n^\alpha$ describing the photons direction in a homogeneous universe and in a global coordinate. The difference becomes subtle at the observer position, as we described in appendix \ref{Identification of the GLC angles with the observed angles}. However, when these two vectors are contracted with perturbation quantities the result at linear order is identical, as the difference in the two vectors appears at perturbative level. As a consequence, we can write the photon wavevector in a global coordinate as
\begin{equation}
		k^\mu_{\text{FRW}} = \frac{\omega}{a}(-1 + \mathcal A +  n^\alpha V_\alpha - n^\alpha \mathcal B_\alpha \,,\,  n^\alpha -  V^\alpha -  n^\beta \mathcal C^\alpha_\beta )\,.
\end{equation}
It should be clear that the above quantity, even though it is expressed in a global coordinate, is physically meaningful only locally at the observer position, where the observer rest frame is defined.

\section{Calculation of the gravitational lensing convergence}
\label{appendix B}

In this appendix we calculate the gravitational lensing convergence $\kappa$ (or $J_2$ in \cite{GLC2012}) when the degrees of freedom in the GLC angles are fixed in such a way that the GLC angles match the observed angles in the observer rest frame (see appendices \ref{Identification of the GLC angles with the observed angles} and \ref{appendix C}). The quantity we have to calculate is  
\begin{equation}
	\kappa \equiv \frac{1}{2}\partial_a \delta\tilde\theta^a + \frac{1}{2}\cot \tilde\theta \, \delta \tilde\theta = \frac{1}{2}\hat\nabla_a \, \delta\tilde\theta^a \,.
\end{equation}
To simplify this task, we make use of three unit directional vectors: $n^\alpha$, $\vartheta^\alpha$, $\varphi^\alpha$, orthogonal to each other. 
The observed angular position of the source is represented by the unit vector\footnote{In this appendix we drop the subscript ``obs'' to refer to the observed angles.}
\begin{equation}
\label{n}
n^\alpha=(\sin\theta\cos\phi,\sin\theta\sin\phi,\cos\theta)	\,.
\end{equation}
Based on $n^\alpha$, we define two unit vectors generating the tangent plane to the two-sphere parametrized by $(\theta , \phi)$ at the point where $n^\alpha$ is attached:
\begin{equation}
\label{t p}
\begin{split}
& \vartheta^\alpha = \partial_\theta n^\alpha = (\cos\theta\cos\phi , \cos\theta\sin\theta , -\sin\theta) \,,
\\
& \varphi^\alpha = \frac{1}{\sin\theta}\partial_\phi n^\alpha = (-\sin\phi , \cos\phi , 0) \,.
\end{split}
\end{equation}
In spherical coordinates these unit vectors are $n_\alpha=(1,0,0)$, $\vartheta_\alpha=(0,r,0)$, $\varphi_\alpha=(0,0,r\sin\theta)$,
and their product with a generic spatial vector $A^\alpha$ gives respectively the radial component and the two angular components:
\begin{equation}
n_\alpha A^\alpha = A^r \,, \qquad\qquad \vartheta_\alpha A^\alpha=r A^\theta \,, \qquad\qquad \varphi_\alpha A^\alpha=r\sin\theta \, A^\phi \,.	
\end{equation}
Consequently, starting from the expression of a given quantity in spherical coordinates, we can rewrite it in a covariant way by using the unit vectors. 
After that, we can make use of any coordinate system to perform the calculations. 
Indeed, the calculation of $\kappa$ greatly simplifies if we first rewrite $\delta \tilde \theta^a$ given by eq.~\eqref{final delta theta} as
\begin{equation}
	\begin{split}
		\delta\tilde\theta &= -\theta_\alpha \big[-\mathcal V^\alpha + \Psi^\alpha + C^{[\alpha|\beta]}n_\beta + C^{\alpha\beta} n_\beta \big]_o  +\frac{\theta_\alpha \mathcal G^\alpha}{\bar r_z}\bigg\vert^s_o 
		\\
		&  +\int_0^{\bar r_z}d\bar r\,\bigg[ \frac{\theta_\alpha (\Psi^\alpha + 2\,C^\alpha_\beta \, n^\beta )  }{\bar r} 
		  + \bigg(\frac{\bar r_z - \bar r}{\bar r_z \bar r}\bigg) \partial_\theta(\alpha_\chi -\varphi_\chi - \Psi_\beta \, n^\beta - C_{\beta\gamma}\, n^\beta n^\gamma )\bigg]		\,,
	\\
		\delta\tilde\phi &=- \frac{1}{\sin\theta} \phi_\alpha \big[-\mathcal V^\alpha + \Psi^\alpha + C^{[\alpha|\beta]}n_\beta + C^{\alpha\beta} n_\beta \big]_o  +\frac{\phi_\alpha \mathcal G^\alpha}{\bar r_z \sin\theta}\bigg\vert^s_o 
		\\
		&   +\int_0^{\bar r_z}d\bar r\,\bigg[ \frac{\phi_\alpha (\Psi^\alpha + 2\,C^\alpha_\beta \, n^\beta )  }{\bar r \sin\theta} 
		  + \bigg(\frac{\bar r_z - \bar r}{\bar r_z \bar r}\bigg) \frac{1}{\sin^2\theta} \partial_\phi(\alpha_\chi -\varphi_\chi - \Psi_\beta \, n^\beta - C_{\beta\gamma}\, n^\beta n^\gamma )\bigg]	\,,
	\end{split}
\end{equation}
and we choose cartesian coordinates, so that any covariant derivative with respect to the three-spatial metric $\bar g_{\alpha\beta}$ reduces to an ordinary derivative, as $\bar g_{\alpha\beta}=\delta_{\alpha\beta}$. 
After introducing the angular gradient and the angular Laplacian, 
\begin{equation}
	\hat\nabla_\alpha = \theta_\alpha\, \partial_\theta  +  \frac{1}{\sin\theta} \phi_\alpha \, \partial_\phi \,, \qquad \qquad \hat\nabla^2 = \partial_\theta^2+\cot\theta\,\partial_\theta+\frac{1}{\sin^2\theta}\partial_\phi^2 \,,
\end{equation}
and noting the identity 
\begin{equation}
	(\cot\theta + \partial_\theta)\,\theta_\alpha  +  \frac{1}{\sin\theta}\partial_\phi \phi_\alpha  = -2 n_\alpha \,,
\end{equation} 
we derive the gravitational lensing convergence
\begin{equation}
	\begin{split}
		\kappa &=  n_\alpha \big[-\mathcal V^\alpha + \Psi^\alpha  + C^\alpha_\beta n^\beta \big]_o + \frac{1}{2\bar r_z} \hat\nabla_\alpha \mathcal G^\alpha - \frac{n_\alpha \mathcal G^\alpha}{\bar r_z} \bigg\vert^s_o - \int_0^{\bar r_z}d\bar r\, \frac{n_\alpha (\Psi^\alpha + 2\,C^\alpha_\beta n^\beta )}{\bar r}
		\\
		 & \quad  
		 + \frac{1}{2} \int_0^{\bar r_z}d\bar r\, \bigg[ \frac{\hat\nabla_\alpha  (\Psi^\alpha + 2\,C^\alpha_\beta n^\beta )}{\bar r} 
		 + \bigg(\frac{\bar r_z - \bar r}{\bar r_z \bar r}\bigg) \hat\nabla^2 (\alpha_\chi -\varphi_\chi - \Psi_\alpha n^\alpha  - C_{\alpha\beta}n^\alpha n^\beta )\bigg] \,.
	\end{split}
\end{equation}
Finally, going back to spherical coordinates, we obtain:
\begin{equation}\label{J2}
\begin{split}
\kappa  =&  [-\mathcal V^r + \Psi^r + C^{rr}]_o  + \frac{1}{2}\hat\nabla_a  \mathcal G^a  + \frac{1}{\bar r_z}\mathcal G^r_o
\\
&+ \frac{1}{2}  \int_0^{\bar r_z}d\bar r \bigg[ \hat\nabla_a \big(\Psi^a +2\,C^{ra}\big) + \bigg(\frac{\bar r_z-\bar r}{\bar r_z \bar r}\bigg) \hat\nabla^2 (\alpha_\chi - \varphi_\chi -\Psi^r - C^{rr}) \bigg] \,,
\end{split}
\end{equation}
where $\hat\nabla_a \Psi^a=\partial_a \Psi^a + \cot\theta \, \Psi^\theta$. This result is probably the most complicated to derive but is in agreement with the gravitational lensing convergence calculated in \cite{Yoo} with the geometric approach and with fully general metric representation to first order in perturbations.

\end{document}